\documentclass{aa501}
\usepackage{graphicx}
\newcommand{\al}{et al.}
\newcommand{\hi}{H{\sc i}}
\newcommand{\prim}{$^{\prime}$}
\newcommand{\prin}{$^{\prime\prime}$}
\newcommand{\aprox}{${\sim}$}

\newcommand{\km}{km\,s$^{-1}$}
\newcommand{\degree}{$^{\circ}$}
\newcommand{\halpha}{H${\alpha}$}
\newcommand{\cg}{CGCG\,160--}
\newcommand{\msolar}{M$_{\odot}$}

\newcommand{\mhi}{M$_{\mathrm {HI}}$}
\newcommand{\mhdos}{M$_{\mathrm {H2}}$}

\newcommand{\vrel}{v$_{\mathrm {rel}}$}
\newcommand{\rstr}{R$_{\mathrm {str}}$}
\newcommand{\angs}{${\mathrm {\AA}}$}
\newcommand{\msolaryr}{M$_{\odot}$\,yr$^{-1}$}

\begin{document}
   \title{VLA HI Imaging of the brightest spiral galaxies in Coma}

   \subtitle{II. The HI Atlas and deep continuum imaging of selected early type
  galaxies }

   \author{H. Bravo--Alfaro
          \inst{1},
V. Cayatte \inst{2}, J. H. van Gorkom \inst{3}, and C. Balkowski  \inst{2}  
          }

   \offprints{hector@astro.ugto.mx}

   \institute{\inst{1} Departamento de Astronom\1a, Universidad de Guanajuato.
Apdo. Postal 144, Guanajuato 36000. M\'exico\\
   \inst{2} Observatoire de Paris DAEC, and UMR 8631, 
associ\'e au CNRS et \`a l'Universit\'e Paris 7, 92195 Meudon Cedex, France \\
   \inst{3} Department of Astronomy, Columbia University, 550 W 120th Street,
New York, NY 10027 USA 
             }

   \date{Received ; accepted }

   \abstract{
   In the first paper of this series we used \hi~observations of the 19
brightest spirals in Coma to analyze the dynamical state of the cluster. In
this paper we present the detailed \hi~distribution and kinematics of the
spirals that were detected in \hi, and radio continuum data for a sample of
star forming and post starburst galaxies in Coma. We discuss the importance
of ICM-ISM interactions to explain the observed \hi~morphology.  A rough
comparison of observed \hi~sizes with predicted \hi~sizes from simulations by
Abadi et al.\,(1999) gives reasonable agreement.  We use the results on radio
continuum emission to estimate the star formation rate in the PSB galaxies we
pointed at. The radio continuum emission in the 11 so called post starburst
galaxies, identified by Caldwell et al.\,(1993) in the cluster, is
weak. Eight of the 11 were not detected down to a 3 sigma upper limit of
0.6\,mJy. This sets an upper limit to the star formation rate in these
galaxies of less than 0.2\,\msolar\,yr$^{-1}$. The three detected post
starburst galaxies have a star formation rate of less than one solar
mass per year. Thus none of the post starburst galaxies in Coma are dust
enshrouded starbursts.
   }

   \maketitle
%

\section{Introduction}

This is our second paper on the imaging of the neutral hydrogen component as
a tracer of environmental effects on galaxies in the Coma cluster. Global
properties of Coma related with its dynamical state, as derived from our VLA
21\,cm line survey, have been discussed in a previous paper (Bravo--Alfaro
\al\,2000, hereafter referred to as Paper\,I). In the present work we give a
complete catalogue of \hi~maps, channel maps, and velocity fields (for cases
with sufficient resolution) for 19 detected galaxies.  We compare the
detailed \hi~distribution and kinematics with model predictions for the fate
of \hi~in galaxies in a cluster environment. We pay particular attention to a
class of so called starburst (SB) and post starburst (PSB) galaxies
identified by Caldwell et al.\,1993 (C93 throughout this paper).  For these
galaxies we derive star formation rates (SFR) from the radio continuum
emission and discuss the implications of current gas content and star
formation rate on the evolution of the galaxies.

Several physical mechanisms have been proposed to play a major role affecting
the evolution of galaxies in clusters; these processes can be classified
according to three types: (a) interactions between the ICM and the
interstellar medium (Gunn and Gott 1972, Nulsen 1982, Kenney and Young 1989),
(b) interactions with the cluster gravitational field (e.g. Bekki 1999), and
(c) galaxy--galaxy interactions (see Icke 1985, Lavery and Henry 1994, Moore
\al\,1996, 1999). In many cases more than one of these processes will be at
work, but which mechanism dominates under specific physical conditions and
how it depends on the galaxy trajectory through the cluster is still a matter
of debate. For instance Solanes et al.\,(2000) conclude from an observational
study of eighteen clusters that ICM-ISM interactions are most important to
explain the observed \hi~deficiency. In contrast, Moore et al.\,(1999) suggest
in a numerical study that, in combination with ram--pressure stripping,
galaxy harassment may convert disk galaxies into S0s. In spite of this
debate, the fact that spirals in the central region of clusters like Virgo
show smaller \hi~than optical disks (Warmels 1988, Cayatte \al\,1990, 1994),
is unquestionably associated with the interaction with the ICM. These studies
demonstrate the importance of detailed \hi~imaging of individual galaxies in
clusters with a range of ICM conditions to assess what gas removal processes
are at work.

Another fundamental process linked with the evolution of galaxies in clusters
is the triggering and stopping of bursts of star formation (van den Bergh
\al\,1990). As the ISM is the raw material from which stars form, the
evolution of \hi~deficient galaxies and their star formation properties are
undoubtedly affected by the interaction with the environment. From
statistical analyses on large scales the SFR is known to decrease with
increasing density (Dressler \al\,1999), whereas at smaller scales the ICM
may produce significant enhancements to the star formation activity in
individual galaxies, as shown in clusters like A\,1367 (Gavazzi \al\,1995)
and Coma (Bothun and Dressler, 1986). Recent work supports different
scenarios: (1) On the theoretical side, Bekki (1999) concludes that the tidal
gravitational field of a group--cluster merger may trigger a burst of star
formation, accounting for the PSB population in clusters like Coma, while
Vollmer \al~(2001b) suggest that local enhancement of star formation could be
due to re-accretion of gas clouds after a ram pressure stripping event. (2)
Observationally, Dressler \al\,(1999) find a trend in the position of SB and
PSB galaxies around the cluster, making the hot gas environment the best
explanation for this phenomenon (see also Poggianti \al\,1999).  From a study
of spectroscopically selected post starburst galaxies, which turned out to be
mostly located in the field, Zabludoff et al.\,(1996) conclude that in those
more isolated galaxies the post starburst phenomenon is most likely caused by
galaxy interactions and mergers.

In this paper we give the atlas of our VLA \hi~imaging of spirals in
Coma, showing that the scenario depicted by the neutral hydrogen
constitutes strong evidence in favor of ICM-ISM interaction to explain
the \hi~morphologies and the star formation histories in the
cluster. We also present results on the radio continuum emission
obtained as a byproduct of the spectral line observations. These
results are used to derive the star formation rate in the so called
post starburst galaxies to explore the possibility that these are in
fact dust enshrouded starbursts as has recently been suggested by
Smail et al (1999) for the intermediate redshift cluster CL0939+4713
(see also Miller \& Owen, 2001).

We organize the paper as follows. In Section\,2 we review the
observations. In Section\,3 we describe \hi~observational results of
individual detected galaxies, in Section\,4 we discuss the role of ram
pressure affecting galaxies in Coma, and compare \hi~observations with
numerical simulations of ram pressure stripping by Abadi \al\,(1999) and
Vollmer \al\,(2001b). In Section\,5 we discuss the PSB galaxies; we compare
our \hi~mass limits and current star formation rates based on the continuum
with those of the more vigorously star forming galaxies and discuss the
implications. The results are summarized in Section\,6. In Appendix~A,
available in electronic form, we give the complete catalogue of \hi~channel
maps.  Throughout the paper we assume a distance of 70\,Mpc to the Coma
cluster (with H$_0$=100\,km\,s$^{-1}$\,Mpc$^{-1}$), where an angular size of 1
arcmin corresponds to a linear size of \aprox20\,kpc.

\section{Observational results}

Our data consist of 21\,cm line data obtained with the VLA\footnote{The
National Radio Astronomy Observatory is operated by Associated Universities,
Inc., under cooperative agreement with the National Science
Foundation.}. Twelve fields within one Abell radius of the Coma cluster
(equivalent to 1.2\degree) were observed with the VLA in its C configuration,
and two of them reobserved in D configuration. The observed fields and the
distribution of the 19 galaxies detected in \hi~around the cluster are shown
in Paper\,I (Fig.\,1 and Fig.\,2). Most of the fields are devoted to the
center of Coma and the regions where SB and PSB galaxies had been reported
from optical observations. Our velocity resolution is 21\,\km, and
43\,\km~for some of the galaxies observed in both C and D configurations. The
angular resolution ranges between 20 and 35\,arcsec. The rms in the final
cubes is between 0.35 and 0.40 mJy beam$^{-1}$, except in those fields
observed in both C and D configurations, reaching rms values as low as 0.20
mJy beam$^{-1}$. Most of the channel maps shown in the Appendix (electronic
version) are smoothed in velocity. More details on the \hi~observations are
given in Paper\,I.

Eight out of the 19 detected galaxies are projected within 0.5\,Mpc from the
cluster center (we consider the position of the cD NGC\,4874 as the Coma
center). These central galaxies lie inside or near the hot ICM as traced by
the ROSAT X-ray emission (Briel \al\,1992 and Vikhlinin \al\,1997), and most
of them are very \hi~deficient. The \hi~deficiency is measured following
Giovanelli \& Haynes (1985), comparing the observed \hi~mass with the
expected \hi~mass of an equivalent isolated spiral.  Several galaxies
detected in \hi~near the center of Coma show \hi~truncated disks as well as
offsets between the optical and \hi~distributions (see Sect.\,3). In a few
cases unexpected differences between optical and \hi~velocities are
found. Six of the galaxies detected in \hi~are reported by Bothun and
Dressler (1986) as having blue disks; five of these are projected inside or
near the cluster X-ray emission. None of the PSB galaxies reported by Caldwell
\al\,(1993, 1997) were detected in \hi; we obtain for some of them \hi~mass
upper limits down to 3$\times$10$^7\,$\msolar.

Our radio continuum images are obtained as a byproduct of the
\hi~observations. The images are built by averaging a set of those line free
channels in a given cube. All of the \hi~detected spirals (except NGC\,4907)
are also detected in the continuum with flux densities ranging between 2 and
16\,mJy (see Paper I).

\section{Results for \hi~detected galaxies}

In this section we briefly describe the \hi~morphology of detected galaxies,
and the remarkable features concerning galaxy dynamics as revealed by the
\hi.  \\


\noindent
{\bf IC\,3913, KUG\,1255+275, and Mrk\,057}

These three galaxies were detected in the SW of Coma, between 17\prim~and
25\prim~south of the cD galaxy NGC\,4839.  They display several common
features: they are \hi~rich galaxies with regular gas distributions, where
\hi~disks extend well beyond the optical (Figs.\,1, 3 and 4). Systemic
velocities for these galaxies are between 7400\,\km\,and 7650\,\km, close to
the value of the NGC\,4839 group, 7339\,\km~(Colless \& Dunn, 1996). Because
of their position, lying far from the cluster center and outside the X-ray
emission, no strong interaction with the ICM is expected. Although these
three galaxies are likely to be part of the infalling NGC\,4839 group
(Neumann \al~2001), we see no signs of a gravitational interaction either
optically or in \hi~or the radio emission. Velocity fields of IC\,3913 and
Mrk\,057 are shown in Figs.\,2 and 5 respectively.\\

\noindent
{\bf NGC\,4848}

This Scd galaxy is one of the most interesting objects in our sample.
NGC\,4848 is a blue disk galaxy showing a very intriguing \hi~distribution
(see Fig.\,6), as we only detect gas in the northern edge of the galaxy,
19\prin~(6.5\,kpc) NW from the optical center. We are probably missing some
\hi~flux previously detected by single dish observations in the center of the galaxy (Gavazzi, 1989). This object was also observed in \halpha~by Amram
\al\,(1992) who reported a double peak in the \halpha~profile, with the
northern \halpha~peak at the same position than the \hi.  Interestingly, the
brightest HII regions are in the south, where the \hi~gas has been totally
depleted. The star forming activity is also evidenced by 20\,cm radio
continuum emission above a level of 16\,mJy. This emission shows an
elongation to the NW which is resolved by the VLA FIRST Survey in two
separated sources, one coincident with the optical center of NGC\,4848 and a
secondary peak 12\prin~NW, roughly at the same position than the \hi. The CO
imaging carried out by Vollmer \al\,(2001a) shows the maximum of the emission
coincident with the optical galaxy center and with the empty \hi~zone. A
secondary CO peak is detected in the NW, roughly coincident with the \hi~and
radio continuum emission.  Lavezzi \al\,(1999) reported a normal
\mhdos~content (2.56$\times$10$^9$\,\msolar), which gives a very low fraction
of neutral to molecular gas ratio: \mhi/\mhdos\aprox 0.17 (NGC\,4848 is
\hi~deficient by a factor of around 10).

Mergers do not account for the \hi~morphology or the star formation activity
in NGC\,4848, because no obvious companion is seen in the DSS optical image
(Fig.\,6). However, the more detailed B-band CCD imaging by Gavazzi
\al~(1990) shows a ring-like structure and blue bright zones in the NW, where
the \hi~and the secondary peaks of CO, \halpha~and 20\,cm radio continuum are
found.  The hypothesis of a dwarf system crossing the NGC~4848 disk is
explored by Vollmer \al\,(2001a) but further observations, both optical and
higher resolution \hi~imaging, are needed to confirm it. N--body simulations
by the same authors suggest that NGC\,4848 has already gone through the
cluster core, 4$\times$10$^8$ yr ago, and is now moving away from the
cluster. They conclude that re-accretion of some of the stripped gas could
explain the star formation burst. \\

\noindent
{\bf Mrk\,58}

This Sb, blue disk galaxy is projected onto the X--ray emission, some
20\prim~(\aprox400\,kpc) SW of NGC\,4874, in the zone lying between the main
cluster and the SW group. It is gas deficient by a factor of 3. Its \hi~map
(Fig.\,7) displays a considerably asymmetry, with the gas swept from the
NE. The \hi~is observed at an offset position of \aprox12\prin\,(4\,kpc) SW
from the optical disk. The observed \hi~distribution could be explained if we
consider the supersonic velocity of Mrk\,58 relative to the cluster,
1575\,\km~(the sound speed in Coma is estimated around 1460\,\km~by Stevens
\al\,1999); under these conditions the interaction with the ICM may produce
ram pressure stripping to enhance the density behind the galaxy (Stevens
\al\,1999).  \\

\noindent
{\bf \cg058 and \cg076}

These late--type spiral galaxies lie in the nearby northern vicinity of Coma,
outside the X--ray emission. They are \hi~rich galaxies
displaying regular gas distributions (Figs.\,8 and 10). As expected,
considering their position relative to the cluster, no strong environmental
effects are seen: their \hi~disks are larger than the optical, and the
\hi~and optical centroids are coincident. Velocity fields are given in
Figs.\,9 and 11.\\

\noindent
{\bf \cg086}

This Sb, blue disk galaxy, projected onto the SE outskirts of the X--ray
emission drawn by ROSAT, was only marginally detected in the present
survey. The galaxy is not resolved in \hi, as shown in Fig.\,12. We report an
\hi~deficiency by a factor of \aprox3, the \hi~disk appearing truncated in
its periphery, but the low resolution does not allow to get a more conclusive
picture. \\

\noindent
{\bf IC 4040}

This Sdm blue disk galaxy is projected near the very center of Coma, at only
15\prim~(300\,kpc) NE from NGC\,4874. Its \hi~distribution (see Fig.\,13)
shows the \hi~contours compressed in the NW, and an offset in position of
9\prin~(\aprox3\,kpc), with the \hi~emission SE of the optical galaxy.  Our
detection confirms the double peak detection reported by Gavazzi (1989) at
7600\,\km~and 7850\,\km.  The velocity field for this galaxy is shown in
Fig.\,14. \\

\noindent
{\bf NGC\,4907}

This Sb galaxy was only marginally detected in \hi~(Fig.\,15). It is
\hi~deficient by a factor of ten and presents a shrunken \hi~disk (not
resolved by the VLA beam) which is smaller than the optical disk. Two facts
make an ICM--ISM interaction the most likely explanation for this picture:
the high radial velocity relative to the cluster (v$_{\mathrm
{rel}}$=1179\,\km), and the projected location of NGC\,4907, which puts it in a
very high density region, 20\prim~(400\,kpc) NE from the cluster center. \\

\noindent
{\bf KUG\,1258+287, FOCA\,195 and \cg098}

These objects are part of a group dominated by \cg098 (see Biviano
\al\,1996). As all the galaxies in Coma detected in \hi~outside the X-ray
emission, this group shows a normal \hi~content and a regular gas
distribution (Figs.\,16, 18 and 20), with the \hi~more extended than the
optical disk and coinciding with the optical position. Particular features of
these galaxies are their systematic blue m$_{\mathrm {UV}}$--b color (see
Table\,2 and Fig.\,7 in Paper\,I), and their high radial velocity (between
8400\,\km~and 8880\,\km), indicating a fast accretion of the group towards
the main cluster body. Normal rotation patterns are shown by the bluest
objects KUG\,1258+287 and FOCA\,195 (see Figs.\,17 and 19). \hi~emission
equivalent to 2.2$\times$10$^{8}$ \msolar~was detected at 1.4\prim~W of
KUG\,1258+287 but no optical counterpart is observed. \\

\noindent
{\bf NGC\,4911}

This is one of the two brightest spirals in Coma. The \hi~morphology of
NGC\,4911 shows a shrunken disk with two central peaks and a normal surface
density in the central region as predicted by ram--pressure stripping
(Fig.\,22). A shift in position is shown between the \hi~and the optical, the
former lying about 4\,kpc south of the optical component. NGC\,4911 shows a
very high total gas content,
\mhi\,+\,\mhdos\,\aprox\,3.7$\times$10$^9$\,\msolar~(Casoli \al~1996), but
only a small fraction (\mhi/\mhdos=0.34) is in atomic form. Considering the CO
observations by Lavezzi \al~(1999) this fraction is much lower:
\mhi/\mhdos=0.22. The velocity field shows normal rotation of the inner disk
(see Figs.\,23) along an axis oriented roughly SE-NW.

NGC\,4911, projected 20\prim\,SE of NGC\,4874 (\vrel\aprox1000\,\km), is
thought to be the dominant galaxy of a group that recently crossed the
cluster core (Biviano \al\,1996). Vikhlinin \al~(1997) reported an X--ray
filamentary structure in this region, crossing over NGC\,4911 and culminating
at the position of NGC\,4921. This cool spot has recently been confirmed with
XMM (Arnaud et al.\,2001), who consider the possibility of gas stripped from
an infalling group, but not directly produced in the Coma center. They found
that part of the X-ray excess in this zone is due to NGC\,4911.  No other
galaxies of this group were detected in \hi, in support of the stripping
hypothesis. However, the fact that none of the 14 catalogued galaxies
(Biviano et al.\,1996) have been classified as spirals weakens this argument
somewhat. \\

\noindent
{\bf NGC\,4922}

This is the best known case of a merger in Coma, and the only pair detected
in this survey; it is located in the northern most field, 1.4\degree~from the
cluster center. It is very bright in radio (27.5 mJy) and the brightest IR
source in Coma (Mirabel \& Sanders 1988). NGC\,4922 consists of two merging
galaxies, one spiral in the north and one early type galaxy in the south. We
detected \hi~in emission and absorption. The \hi~shown in Fig.\,24, centered
in the spiral galaxy, is strongly attenuated by the absorption. An
\hi~emission equivalent to 5.8$\times$10$^{7}$ \msolar~was
detected south of the merging system, but no optical counterpart is evident
from the DSS image. \\

\noindent
{\bf NGC\,4921}

This is the brightest spiral in Coma, located 24\prim~(\aprox0.5\,Mpc) SE from
NGC\,4874. It roughly coincides with a second order X--ray peak (Vikhlinin
\al\,1997) also detected with XMM by Arnaud et al.\,(2001) and Briel et
al. (2001). NGC\,4921 presents a very peculiar picture in 21\,cm: a double
peaked \hi~disk (Fig.\,25) which is considerably smaller than the optical
one, with a strongly asymmetric distribution, with most of the neutral gas
distributed along the SE spiral arm. The gaseous disk presents a striking
shift in position of \aprox25\prin~(some 8\,kpc) SE from the optical. The
velocity field is shown in Fig.\,26.

NGC\,4921 is perhaps the best example in Coma where several mechanisms are
present simultaneously. (a) The shrunken \hi~disk and the shift between
optical and \hi~positions look like clear signatures of ram--pressure
stripping, which is also supported by the supersonic velocity of the galaxy
relative to the cluster, 1521\,\km. (b) NGC\,4921 shows the largest cross
section in our sample, which could produce, by viscous stripping, a mass loss
rate value up to 20 \msolar\,yr$^{-1}$. Another feature supporting viscous
stripping is the relatively low surface gas density in the central region of
NGC\,4921, previously classified as {\it anemic} by van den Bergh (1976). (c)
The NW zone, where the brightest HII~regions are seen (Amram \al\,1992),
appears depleted of \hi, similar to the case of NGC\,4848. Furthermore,
NGC\,4921 displays a high total gas content,
\mhi\,+\,\mhdos\,=\,2.65\,$\times$\,10$^9$\,\msolar, with only a small
fraction (0.36) in atomic form, suggesting that \hi~is actively converted to
molecular gas. Gas re-accretion in the NW may also be present, triggering the
HII regions along the spiral arm. As this galaxy does not show any optical
distortion, processes involving gravitational effects are unlikely to be
important. \\

\noindent
{\bf IC\,842 and IC\,4088}

These galaxies lie in the far northern region of Coma, along the supercluster
NE filament, and far from the X--ray emission. As expected, they display
normal \hi~content and no effects of interaction with the ICM are seen
(Figs.\,27 and 29). They also display normal rotation patterns (Figs.\,28
and 30).  Three dwarf systems were detected in \hi~around IC\,4088, one of
them ([GMP 83] 1866) lies at only 2\prim~(\aprox40\,kpc) north of the spiral,
but no optical distortions are evident. The dwarfs show velocity dispersions
between 43\,\km~and 173\,\km, and their \hi~masses range between 0.2 and
2.0$\times$10$^9$\,\msolar.  All the galaxies detected in this region,
including the merger NGC\,4922, are likely part of a group falling towards to
the cluster center (Paper\,I). \\

\noindent
{\bf \cg106}

This blue disk galaxy is located in the SE outskirts of the X--ray emission.
The ICM does not exert important effects on this galaxy because of the low
ICM density at the galaxy position, the low value of \vrel, and the small
cross section. \cg106 is not very \hi~poor (it is deficient by a factor of
1.5), and the \hi~morphology, barely resolved in Fig.\,31, only shows a
shrunken disk slightly bigger than the optical one.

We detect a clear shift in the \hi~position, lying 15\prin~(some 5\,kpc) SW
from the optical disk, and an intriguing difference in velocity between the
\hi~and optical: 6876\,\km~and 7188\,\km, respectively. Amram \al\,(1992)
observed this galaxy in \halpha~and reported a velocity of 7100\,\km~and an
extension to the SW, coincident with the position of a dwarf companion
separated by \aprox20\prin. We detect weak \hi~emission in the same zone, but
higher resolution is needed to resolve the dwarf system.


\section{The \hi~and the ICM in Coma}

As mentioned earlier, physical mechanisms affecting galaxies in clusters are
produced by interactions with one or more of the next three elements: the hot
ICM, neighbor galaxies, and the cluster gravitational field. Two effects
observed in \hi~in this work are explained on the basis of ICM-ISM
interactions: the position of \hi~deficient galaxies relative to the ICM as
drawn by the X--ray emission, and the fact that the most central galaxies
detected in \hi~appear stripped in their outer regions, as predicted for
ram--pressure stripping. If tidal interactions were predominant there should
be evidence of optical signatures of the interaction, but none of the
\hi~deficient galaxies in this work (except the merger NGC\,4922), neither
detected nor undetected, display peculiarities in their optical
morphology. It was also shown in Paper\,I that tidal interactions are unlikely
to be the explanation for the disturbed \hi~disks in Coma or the triggering
of starburst events, because there is no correlation between these effects
and the number of close neighbors.

In order to confirm the role of ram pressure stripping in producing the
observed \hi~distributions and in enhancing the star formation in Coma, it is
fundamental to compare the \hi~imaging with 3D simulations of the ISM--ICM
interaction. A thorough comparison will help to estimate the role played by
different infalling orbits, inclination angles, and ICM densities, as well as
to determine the stripping time scale and the regions of the
disk where ram--pressure is more effective.  To this aim we compare our
\hi~observations with predictions made by 3D-simulations carried out by Abadi
\al~(1999) and Vollmer \al\,(2001b).

Simulations by Abadi \al~(1999) are a good point of departure even if the
clumpiness of the ISM and different elapsed times after crossing the cluster
core are not taken into account. They computed the expected radius of a ram
pressure stripped gas disk (\rstr) for a Coma cluster-like density and
dispersion velocity. For instance, a typical galaxy in Coma with
\vrel\aprox1000\,\km~would have \rstr~\aprox~6\,kpc. In Table\,1 we compare
the observed \hi~radius (or upper limits if resolution is marginal) with the
prediction of R$_{\mathrm {str}}$ made by Abadi \al~(1999). Columns 1 and 2
give the galaxy identification, Column\,3 gives the radial velocity relative
to the cluster. In Column (4) we give the observed \hi~radius in kpc
estimated at a level of 3$\times$10$^{19}$\,\msolar, taking the average
between major and minor \hi~axis (none of the galaxies in Table\,1 are very
elongated). Column\,5 gives the observed \hi~radii corrected for the beam;
for galaxies marginally resolved (indicated with~*) we give 0.5 times the
beam size as an upper limit of the beam corrections, following Wild
1970. Column\,6 gives the predicted value of R$_{\mathrm {str}}$ from Abadi
\al~(1999), considering the corresponding velocity of the galaxy relative to
the cluster.

   \begin{table}
      \caption[]{Comparison between the observed and predicted \hi~radius
      in kpc, for central galaxies in Coma} 
         \label{tbl-1}
     $$ 
	    \begin{tabular}{clrrlrr} 		
            \hline
            \noalign{\smallskip}
CGCG 	& Other 	& \vrel & Obs		& Corr  	& Pred    \\
	& name  	& \km 	& r$_{\mathrm {HI}}$ & r$_{\mathrm {HI}}$
            & r$_{\mathrm {HI}}$  \\
(1) 	& (2)		& (3)	& (4)		& (5)		& (6)	\\ 
\noalign{\smallskip}
\hline
\noalign{\smallskip}
160-055 & NGC\,4848 	& 41  &  6.8 	& ~~5.0*&  17.0  \\
160-073 & Mrk\,058   	& 1575&  7.9 	& ~~5.8	&   4.5  \\
160-086 &          	& 481 &  6.0 	& ~~5.6*&   8.5  \\
160-252 & IC\,4040  	& 758 & 10.2 	& ~~8.0	&   7.5  \\
160-257 & NGC\,4907 	& 1180&  6.6	& ~~5.8*&   5.5  \\
160-260 & NGC\,4911 	& 997 &  12.7 	&12.2	&   6.5  \\
160-095 & NGC\,4921 	& 1521&  11.2 	&10.2	&   5.0  \\ 
160-106 & NGC\,4926-A	& 124 &  8.2 	& ~~5.7	&  14.0  \\  
            \noalign{\smallskip}
            \hline
  \end{tabular}
     $$
(*) galaxies marginally resolved in \hi.
   \end{table}

The predicted \hi~radius is calculated assuming that the galaxy's
distance to the center equals its projected distance and its velocity
through the cluster equals its radial velocity.  Considering those
assumptions there is good agreement between observed and predicted
values of the HI radius. The two galaxies (NGC\,4848 and 4926-A) that
have significantly smaller radii than predicted must have a non
negligible velocity in the plane of the sky, while the giants NGC
4911 and 4921 are probably at larger distance from the center and only
in projection very close. For the three unresolved galaxies the
\hi~distribution is limited to a central region of $\leq$6 kpc.  Our
results for \hi~deficient yet detected galaxies in Coma, confirm that
most of the restoring force is coming from the central parts of the
disk where the presence of the bulge is more important, as found by
Abadi et al.'s simulation. They found that the gas is not completely
removed by ram pressure, suggesting that other processes may be at
work affecting those spirals which are not detected in \hi~in this
survey.

Three dimensional simulations taking into account the clumpiness of the ISM
were carried out for the Virgo cluster by Vollmer et al.\,(2001b). These
authors found that stripping is very effective for galaxies that are on
radial orbits through the cluster, in agreement with observational evidence
provided by Dressler (1984) and Solanes et al.\,(2000).  Vollmer
\al~(2001b) found that time scales for ram pressure effects in Virgo may be
as short as 3$\times$10$^7$\,yr, and in Coma they would likely be shorter
because of its larger core size and more hostile ICM conditions.
Interestingly, those authors found that galaxies showing important gas
disruptions are not infalling but have already gone through the cluster
core. If this is valid in Coma, it would confirm that all the central spirals
(see Fig.\,2 of Paper\,I) have already gone through the cluster core, while
the
\hi~regular galaxies in the outskirts of the X-ray emission are in the
process of infalling (see Sect.\,3). Some of the consequences in Coma are
discussed in the next section.

\section{\hi~and radio continuum of PSB and active galaxies }

The physical mechanisms which trigger and quench the star formation activity,
and the question if the star burst exhausts the \hi~reservoir is still a
matter of debate. We address this problem pointing at 14 galaxies in Coma for
which abnormal spectra are reported by Caldwell \al~(1993, 1997), in order to
get \hi~and radio continuum information. The radio continuum at 1.4\,GHz could
be due to synchrotron emission from relativistic electrons accelerated by
supernovae hence probing star formation, or it could come from AGN. At
intermediate redshift Smail et al.\,(1999) detect in the cluster CL\,0939+4713
(z=0.41) several post starburst galaxies in radio continuum and these authors
argue that the PSBs are most likely dust enshrouded starbursts. In the
local universe  Chang et al.\,(2001) observed in \hi~five of the
spectroscopically selected sample of E+A galaxies (Zabludoff et
al.\,1996). They detected one in \hi~ and none in radio continuum. The
implied low star formation rates and limits on the SFR for these galaxies
rule out that any of these is in fact a dust enshrouded starburst.

Eleven of the 14 abnormal spectrum galaxies we pointed at in Coma are defined
as PSBs because of their spectra with strong absorption lines and no emission
at all; the three remaining are star forming systems (SBs), as indicated by
their emission line spectrum superposed to absorption lines (C93).  Their
position relative to the X-ray emission is shown in Paper\,1 (Fig.\,6). We
detected none of these objects in \hi, and report very low \hi~mass upper
limits between 3 and 7$\times$10$^{7}$\,\msolar~(see Table\,2). We do detect
weak radio continuum emission in three PSB (IC\,3949, Mrk\,060 and RB\,042)
and one SB galaxy (NGC\,4853, see Table\,2).  Our radio continuum levels in
Coma imply even lower limits to the star formation rates than found by Miller
and Owen (2001) and Chang et al.\,(2001) for the field E+A's. Thus in Coma,
contrary to in CL 0939+4713, the post starburst galaxies have at best only
very modest levels of star formation. Chang et al.\,(2001) detected one of
five E+A's in \hi, while our \hi~upper limits for the cluster PSBs are almost
two orders of magnitude below this detection. This may indicate a real
difference between cluster and field post starburst galaxies. Whatever
triggers the starburst, whether collisions or ICM interaction, in a cluster
the gas is cleaned out after the starburst, something that would not happen
in the field. We show that the post starburst galaxies in Coma have very
little, if any, \hi.

   \begin{table*}
      \caption[]{HI and radio continuum parameters of abnormal spectrum
      galaxies (top) and blue disks (bottom) in  Coma}  
         \label{tbl-2}
   $$ 
{\scriptsize
       	    \begin{tabular}{lllcrcccc} 		
\hline 
ID & Other name	& Morph.& \mhi 		& F$_{\mathrm {cont}}$ & rms noise
	& L$_{1.4}$		        &SFR 	& SFR/L$_B$\\

   &		& type	&10$^{8}$\msolar& mJy		       &mJyBeam$^{-1}$
	& $10^{20}$WHz$^{-1}$ & M$_{\odot}$yr$^{-1}$  
& M$_{\odot}$(yr\,L$_{\odot})^{-1}$  \\

(1)&	(2)	& (3)	& (4) 		& (5) 		       & (6)
& (7)		& (8)	& (9)$^a$		      \\ \hline

D 77 & Leda 83676  & S0/a~(SB) &$<$0.6& - & 0.18&$<$3.17& $<$0.19&$<$0.23   \\
D 94 & Leda 83682  & SA0~~(PSB)&$<$0.5& - & 0.18&$<$3.17& $<$0.19&$<$0.17   \\
D 112& Leda 83684  & SB0~~(PSB)&$<$0.4& - & 0.18&$<$3.17& $<$0.19&$<$0.03   \\
D 21 & MCG 5-31-037& SBa~~(PSB)&$<$0.7& - & 0.19&$<$3.35& $<$0.20&$<$0.09   \\
D 73 & RB 183      & SA0~~(PSB)&$<$0.9& - & 0.17&$<$3.00& $<$0.18&$<$0.27   \\
D 44 &KUG1256+278A & S0~~~~(SB)&$<$0.6& - & 0.18&$<$3.17& $<$0.19&$<$0.18   \\
D 43 & NGC~4853    & SA0~~(SB) &$<$0.5&1.2& 0.18&~~7.06& ~~0.42  & ~~0.06   \\
D 89 & IC 3949     & SA0~~(PSB)&$<$0.3&2.1& 0.18&~12.35& ~~0.73  & ~~0.18   \\ 
D 127& RB 042      & S0~~~~(PSB)&$<$0.6&1.0& 0.18&~~5.88& ~~0.35 & ~~0.46   \\
D 216& RB 160      & Sa~~~~(PSB)&$<$0.7& - & 0.18&$<$3.17&$<$0.19&$<$0.28   \\
D 99 & Mrk 060     & SB0~~(PSB)&$<$0.7&1.2& 0.19&~~7.06& ~~0.42  & ~~0.23   \\
D 146& RB 110      & S0~~~~(PSB)&$<$0.3& - & 0.18&$<$3.17&$<$0.19&$<$0.16   \\ 
D 61 & CGCG 160-104& SA0~~(PSB)&$<$0.7& - & 0.10&$<$1.76 &$<$0.10&$<$0.03   \\
D 189& Leda 83763  & S0~~~~(PSB)&$<$1.5& - & 0.17&$<$3.00&$<$0.18&$<$0.31   \\
\hline
160-055 &NGC~4848   & Scd & 4.3 &16.6& 0.18 & ~~97.37 & ~~5.75 & ~~0.38 \\
160-073 &Mrk~058    & Sb  & 2.0 &5.5& 0.17 & ~~32.10  & ~~1.89 & ~~0.56 \\
160-086 &           & Sb  & 1.7 &3.8& 0.15 & ~~22.11  & ~~1.30 & ~~0.88 \\
160-252 &IC~4040    & Sdm & 3.3 &15.0& 0.18 & ~~88.20 & ~~5.20 & ~~1.00 \\
160-098 &           & Sbc & 7.3 &5.7& 0.18 & ~~33.34  & ~~1.97 & ~~0.62 \\
160-106 &NGC~4926--A& Sa  & 6.0 &3.1& 0.15 & ~~18.17  & ~~1.07 & ~~0.49 \\
\hline 
$^a$ in units of 10$^{-9}$.
   \end{tabular}
}
   $$ 
       \end{table*}

Table~2 gives in the upper part the \hi~mass upper limits and radio continuum
parameters for the sample of abnormal spectrum galaxies that we observed. For
comparison, we show in the bottom the same results for the blue
disks. RB\,042 is projected very close to the cD galaxy NGC\,4874 and was only
marginally detected in radio. Three of four galaxies are new detections, the
fourth, NGC\,4853, is catalogued in the FIRST Survey (Becker
\al\,1995). Table\,2 gives in Columns 1 and 2 the galaxy identification, Column
3 gives the morphological and spectral classification, following C93. In
Column\,4 we list the \hi~upper limits, in Column\,5 the radio continuum
flux, Column\,6 lists the rms noise in the continuum images.  In Column\,7 we
give the radio power, with upper limits set at 3 sigma.  The associated SFR,
listed in Column\,8, is calculated following Yun, Reddy \& Condon (2001):
SFR$_{1.4}$\,=\,L$_{1.4}$\,[5.9 $\times$ 10$^{\mathrm
{-22}}$\,W\,Hz$^{-1}$]\,\msolaryr.  In Column 9 we give the relation between
the SFR and the blue luminosity (L$_{\mathrm B}$) obtained from the blue
total magnitude, using the Mean Data from the LEDA database for homogeneity.




Table\,2 shows that the SFR in the PSBs is well below 1\,\msolaryr. It comes
as somewhat of a surprise that also the SB galaxies have SFR less than
1\,\msolaryr. This contrasts sharply with the blue disk galaxies reported by
Bothun and Dressler (1986), which were all detected in continuum (see Table
2) and for which the associated SFR are from 1.0 to 5.5\,\msolaryr. We note
that these rates are in good agreement with the rates reported from the
\halpha~equivalent widths by Bothun and Dressler (1986). Thus the blue
"starforming" disks have a significantly higher SFR than the "starbursting"
galaxies identified by C93. This is also valid if we consider the factor
SFR/L$_B$ given in Column 9 of Table~2, which is probably a better criteria
than merely SFR if the sampled galaxies span a large range in size and
luminosity. There is a consistent trend between Columns~8 and 9, with blue
disks showing systematically higher values of SFR/L$_B$ than galaxies from
C93.  The explanation is simple: C93 only selected early type galaxies for
their sample of abnormal spectrum galaxies. The emission lines typically have
narrow H$\delta$ lines ($\leq$\,3\angs), while the blue disks have
3\,\angs\,$\leq$\,H$\delta$\,$\leq$\,7\,\angs.  Thus the SB in the C93 sample
means a higher than usual SFR for an early type galaxy; the absolute
levels are very low.  It is interesting that the C93 sample is spatially
distinct from the blue disks. The latter are located at the edge of the X
ray emission, while most of the C93 galaxies are in the cluster core or in
the zone between the core and the SW group. The difference in environment may
well account for the different star formation properties of these galaxies.

HST images (Caldwell et al.\,1999) show that some of the PSB galaxies in Coma
have retained their disk, even with a smooth spiral structure.  D61 (Zw
160-104) shows two dust lanes forming an edge-on disk, and D216 (RB 160)
shows clear spiral structure. Interestingly our limits to the SFR and
\hi~content for these two galaxies are very low.  Obviously the time scale
for gas removal from these galaxies was shorter than for a morphological
transformation.  Caldwell \al\,(1999) suggest that star formation activity in
the PSBs of Coma could be triggered by galaxy harassment and by gravitational
perturbations between the main cluster and the SW group. However, Moore
\al~(1996) estimate a time-scale of \aprox3\,Gyr to produce considerable
changes in the optical morphology by this process. This is too long for
galaxies like D\,61 and D\,216 to explain the whole process of triggering a
starburst, the loss of a big fraction of their gas reservoir, and the
subsequent quenching of the activity.  An alternative process to explain this
scenario is the re--accretion of gas mass after the stripping, as the
infalling gas colliding with the clouds remaining in the disk can trigger
star formation within the disk. In a cluster like Virgo this process starts
at \aprox2$\times$10$^8$ years after the closest passage of the cluster core,
and ends at \aprox5$\times$10$^8$ years (Vollmer
\al\,2001b). The same process in Coma should develop with at least the same
time-scales (or shorter, if we consider the more hostile ICM conditions),
suggesting that the interaction with the ICM and gas re--accretion may
account for this scenario. This is reinforced by the fact that galaxies could
reach their present condition during a single pass through the cluster core,
on a time scale $\leq$~1~Gyr, which is much shorter than the time needed by
galaxy harassment to modify the optical morphology. The cases of the PSBs
D\,61 and D\,216 support this scenario: the former presents a very young
starburst age of 0.5\,Gyr, and the latter, recently reclassified as a spiral
(Caldwell \al\,1999), should have harbored a significant amount of gas in the
recent past. Both need a faster process than galaxy harassment to account for
the sudden removal of the \hi~reservoir and quenching of star forming
activity.

The gas which has been accelerated to values below the escape velocity will
accrete back to the galaxy at time scales of 2-7$\times$10$^8$ years
(Vollmer \al\,2001b). As a first approach in Coma we apply a conservative
value of 5$\times$10$^8$ years for the re-accretion time scale, i.e. the same
value found by Vollmer \al~in Virgo. If we consider a typical galaxy moving
across the Coma cluster at \vrel\aprox1000\,\km~(equivalent to the velocity
dispersion) it will travel some 0.5 Mpc before the gas accretes back on to
the disk and triggers a burst of star formation. It is very interesting to
associate this with the fact that a considerable fraction (38\%) of the star
forming UV flux in Coma is produced in an annular region between 20\prim~and
30\prim, or \aprox0.5Mpc from the cluster center (Donas \al\,1995), and that
it is in this zone where most of the blue galaxies in Coma are located
(Paper\,I). Taking into account their position relative to the cluster,
NGC\,4848 and most of the blue disk galaxies in Coma have had enough time
after passing across the cluster core to re-accrete gas clouds and boost a
major star formation event; this mechanism may in part account for the blue
annulus observed in Coma which is also observed in high redshift clusters
(Oemler 1992).

\section{Conclusions}

In this paper we present the \hi~morphology and kinematics of the brightest
spirals in Coma. We compare the \hi~morphology with numerical simulations on
ram pressure stripping by the ICM.  We derive star formation rates for a
sample of post starburst and actively star forming galaxies, from deep
continuum imaging obtained as a byproduct of the \hi~observations.  We
conclude that the \hi~morphology of the spirals in Coma, the location of 
the \hi~deficient galaxies and the size of the \hi~disks are
consistent with predictions of the effect of ram pressure 
stripping by the ICM.

Targeted observations of 11 of the 14 known PSB galaxies in Coma give
\hi~upper limits between 3 and 7$\times$10$^7$\,\msolar~in \hi. The star
formation rates derived from (upper limits to) the radio continuum are less
than 1\,\msolaryr. Even the early type galaxies with abnormal emission lines
(SB galaxies from Caldwell et al.) have SFR well below 1\,\msolaryr. Thus in
Coma there is no evidence for the presence of the dust enshrouded starburst
galaxies, which may have been found in clusters at intermediate redshift. We
found observational evidence suggesting a real difference between cluster and
field post starbursts; galaxies in clusters would exhaust the gas after the
starburst, something that is not always observed in the field.

\begin{acknowledgements}
      
We are very grateful to E. Brinks for helping to improve this paper
significantly. HBA thanks the DAEC of the {\it Observatoire de Paris}, the
Astronomy Department of Columbia University, and the AOC of the NRAO, for
support and hospitality during his visits. We used the Digital Sky Survey,
produced at the Space Telescope Science Institute. We have made use of the
Lyon-Meudon Extragalactic Database (LEDA) supplied by the LEDA team at the
CRAL-Observatoire de Lyon (France). We used NED, the NASA/IPAC extragalactic
database, operated for NASA by the Jet Propulsion Laboratory at Caltech.
This work has in part been supported by NSF grant AST-97-17177 to Columbia
University. We appreciate the suggestions done by an anonymous referee and
the efficiency with which this paper passed through the whole evaluation
procedure.

\end{acknowledgements}

%
   \begin{figure}
   \centering
   \includegraphics[width=7cm]{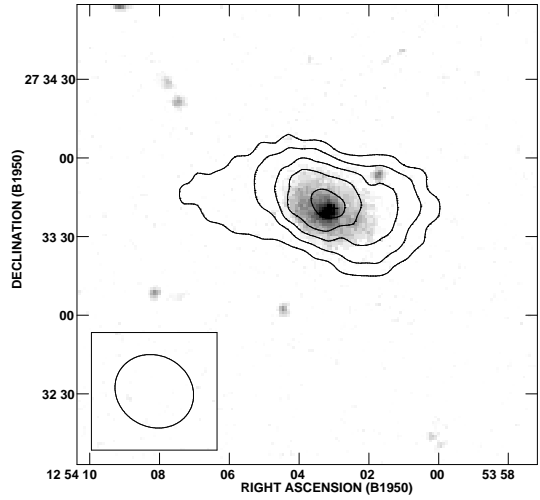}
      \caption{H{\sc i} density distribution of
           IC~3913, superposed on a DSS B-band gray scale image. The contours
           are 0.3 (2.5 $\sigma$), 1.3, 2.7, 4.0, and $5.4 \times 10^{20}$
           cm$^{-2}$. The FWHM is indicated by the circle,
           $30.7^{\prime\prime} \times 27.2 ^{\prime\prime} $.}
\vspace{3cm}
         \label{Fig1a}
   \end{figure}
%

   \begin{figure}
   \centering
   \includegraphics[width=7cm]{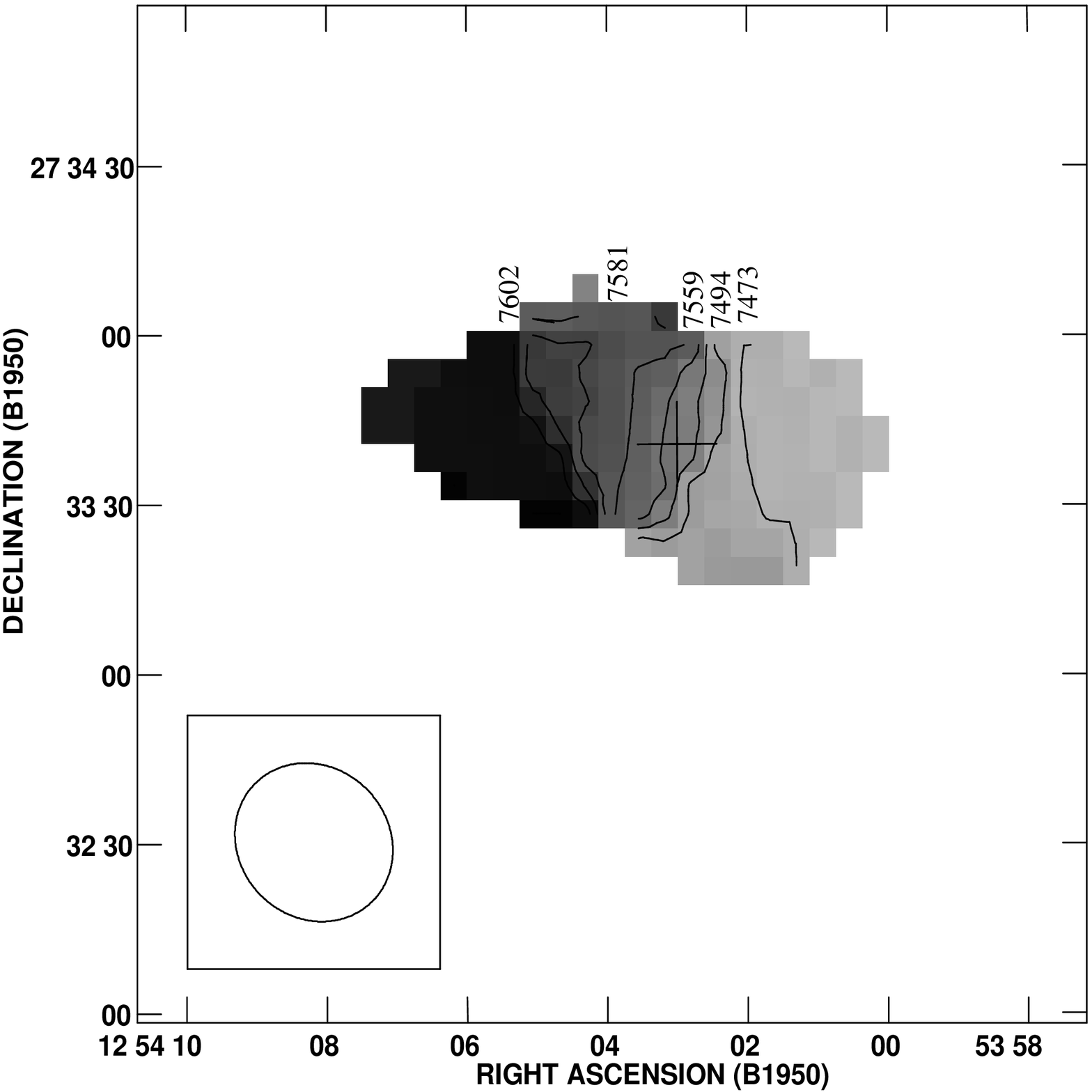}
      \caption{Intensity weighted mean velocity field
           of IC~3913.  The optical center of the galaxy is indicated with a
           cross.  The numbers indicate heliocentric velocity in km
           s$^{-1}$. The FWHM is indicated by the circle, $30.7^
           {\prime\prime} \times 27.2^{\prime\prime}$.
              }
         \label{Fig}
   \end{figure}
%

\newpage

   \begin{figure}
   \centering
   \includegraphics[width=7cm]{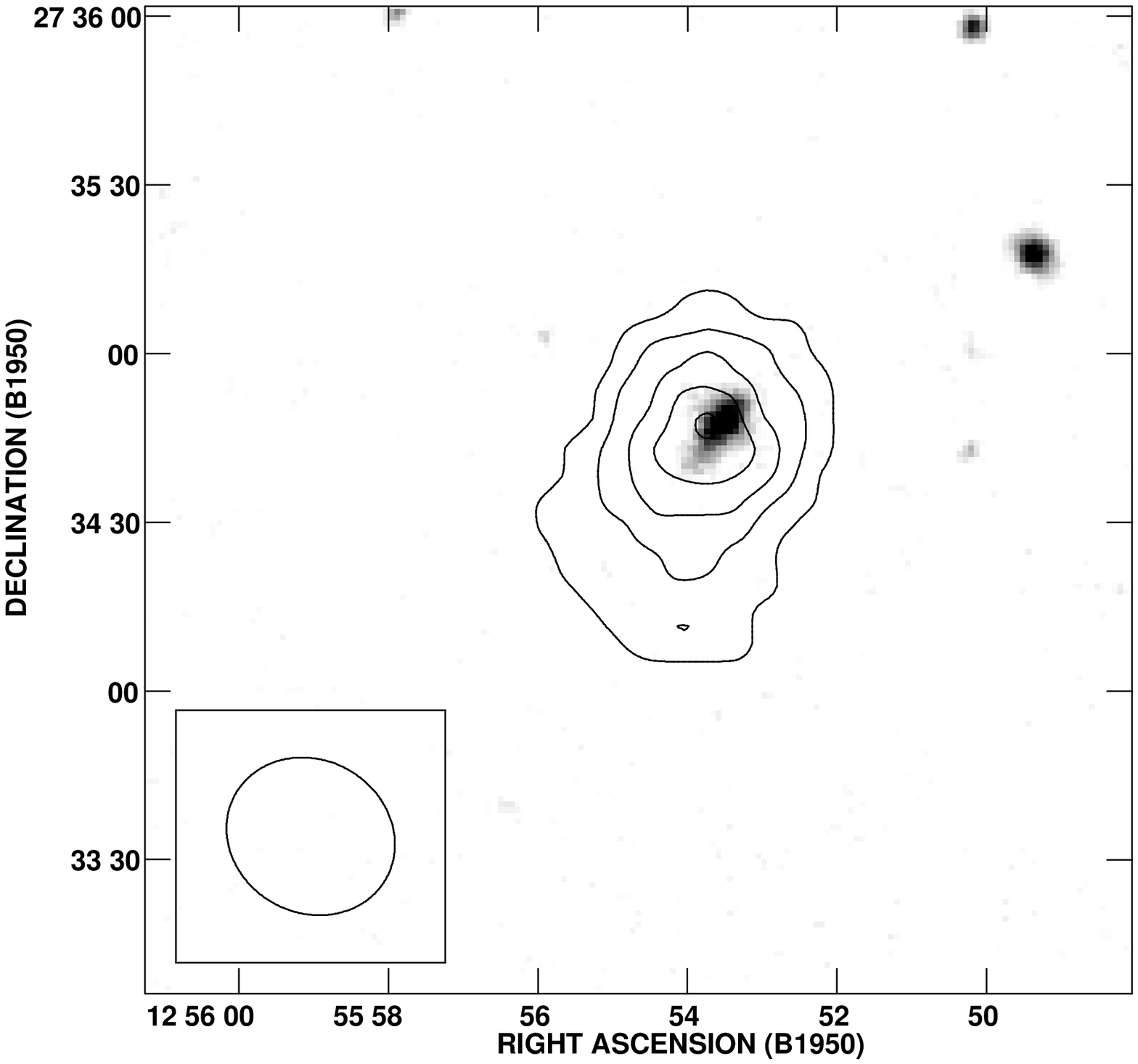}
      \caption{ H{\sc i} density distribution of
        KUG~1255+275, superposed on a DSS B-band gray scale image. The
        contours are 0.3 (2.5 $\sigma$), 0.8, 1.4, 2.0, and $2.6 \times
        10^{20}$ cm$^{-2}$. The FWHM is indicated by the  circle,
        $30.7^{\prime\prime} \times 27.2 ^{\prime\prime} $.
              }
\vspace{15cm}
         \label{Fig}
   \end{figure}
%

\newpage

   \begin{figure}
   \centering
   \includegraphics[width=7cm]{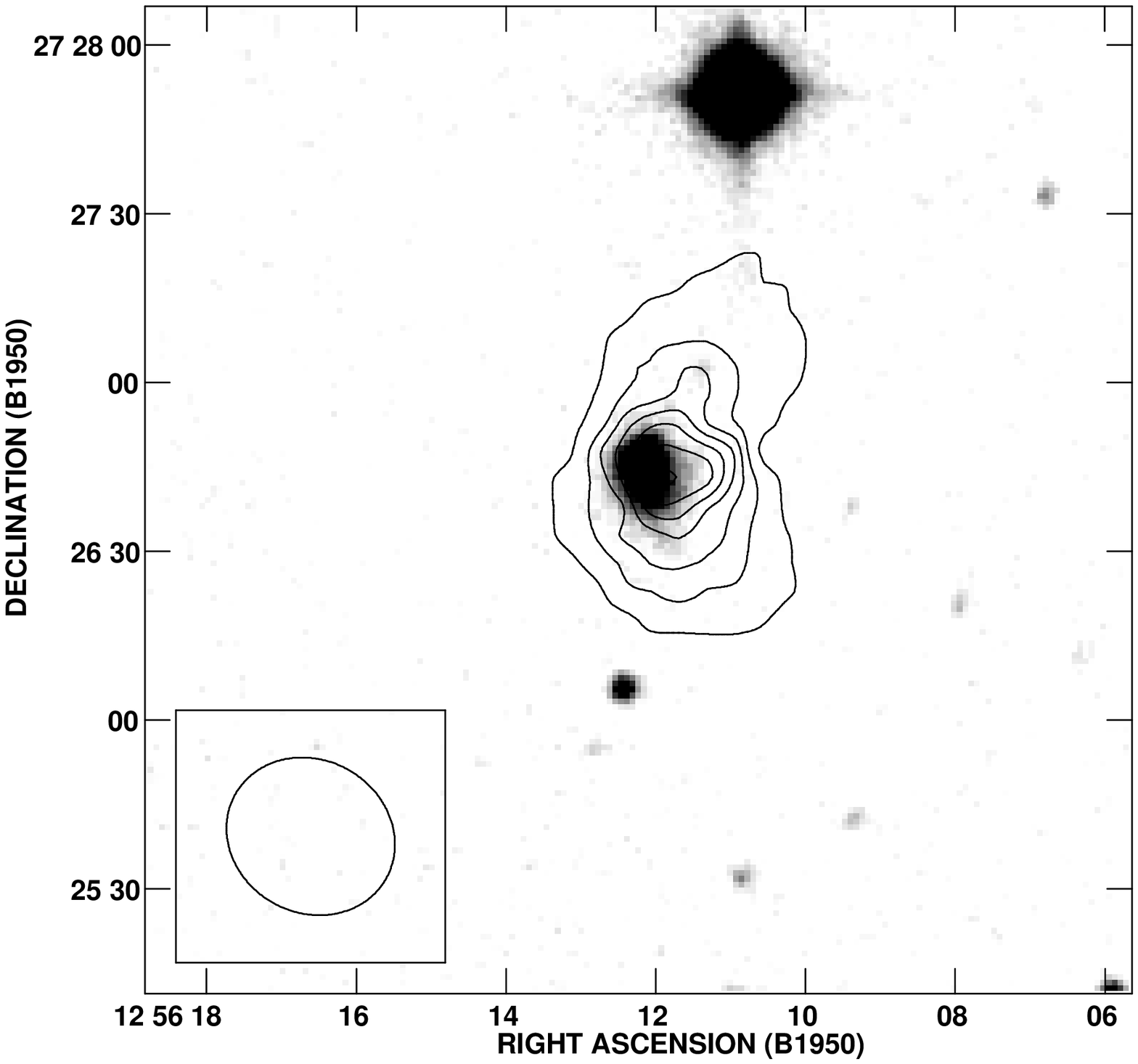}
      \caption{ H{\sc i} density distribution of
           Mrk~057, superposed on a DSS B-band gray scale image. The contours
           are 0.6 (2.5 $\sigma$), 2.3, 3.4, 4.6, 5.7, 6.8 and $8.0 \times
           10^{20}$ cm$^{-2}$. The FWHM is indicated by the circle,
           $30.7^{\prime\prime} \times 27.2 ^{\prime\prime} $.
              }
\vspace{3cm}
         \label{Fig}
   \end{figure}
%

   \begin{figure}
   \centering
   \includegraphics[width=7cm]{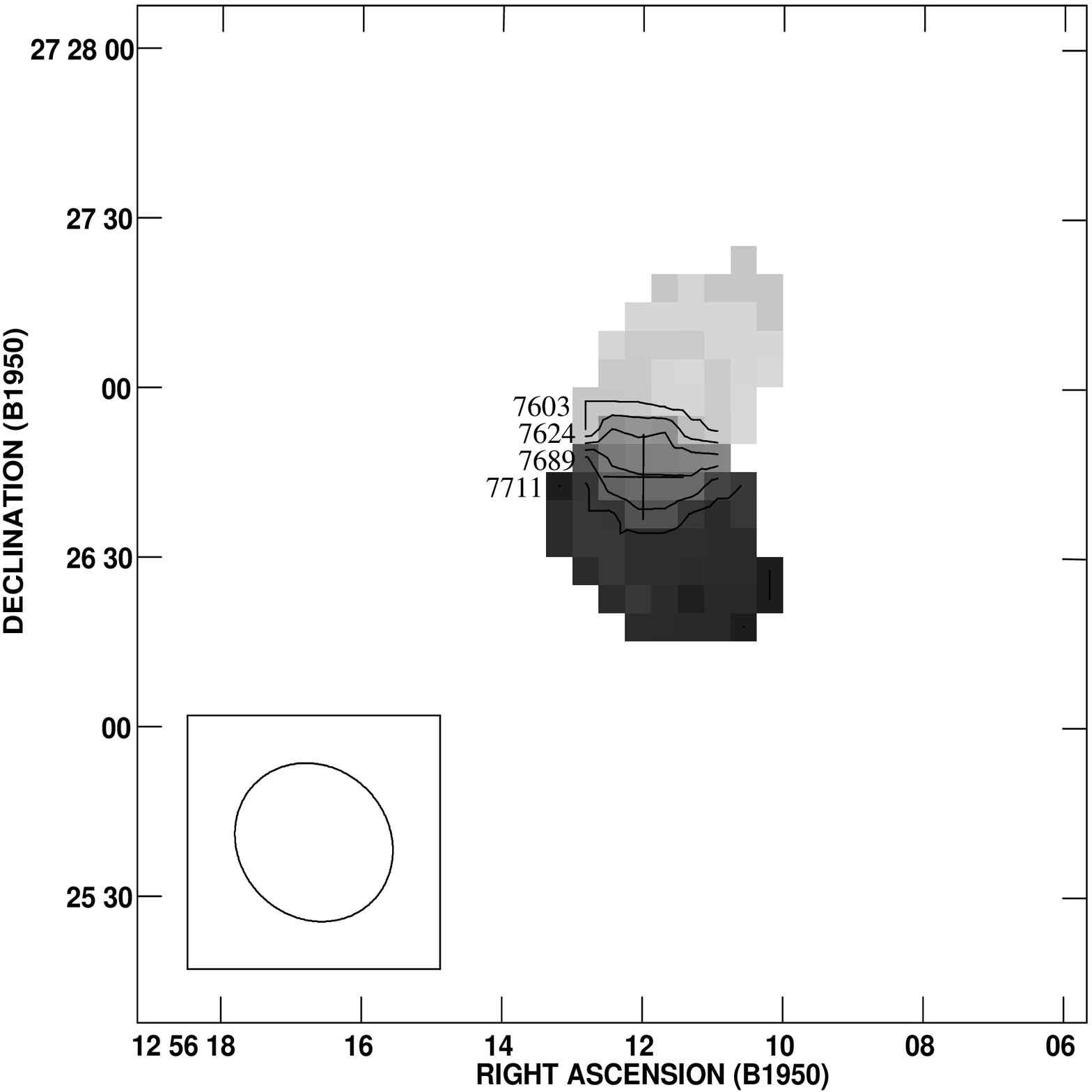}
      \caption{ Intensity weighted mean velocity field
           of Mrk~057.  The optical center of the galaxy is indicated with a
           cross.  The numbers indicate heliocentric velocity in km
           s$^{-1}$. The FWHM is indicated by the circle, $30.7^
           {\prime\prime} \times 27.2^{\prime\prime}$.
              }
         \label{Fig}
   \end{figure}
%

\newpage

   \begin{figure}
   \centering
   \includegraphics[width=7cm]{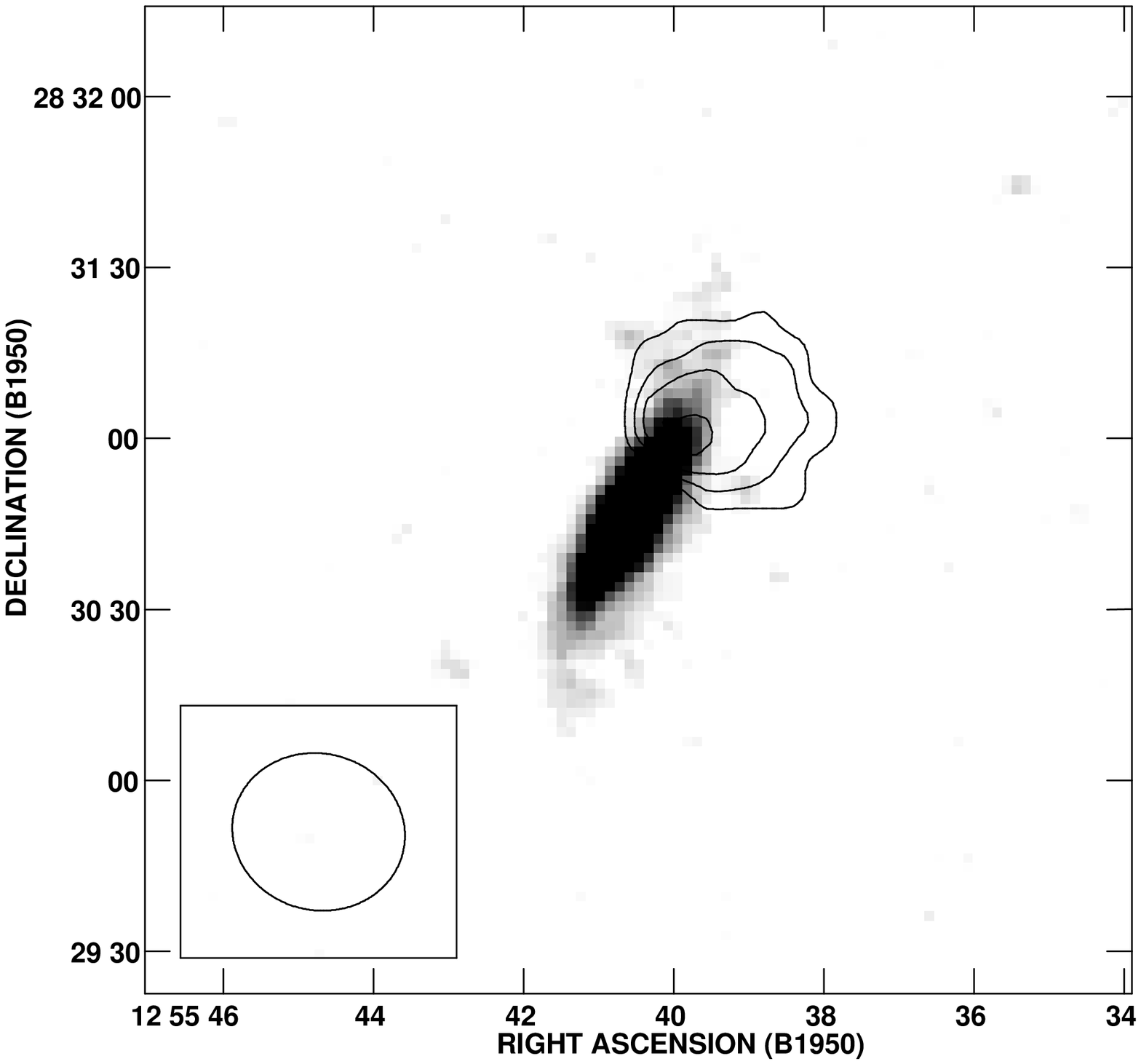}
      \caption{ H{\sc i} density distribution of
           NGC~4848, superposed on a DSS B-band gray scale image. The
           contours are 0.3 (2.5 $\sigma$), 1.0, 1.7, and $2.3 \times
           10^{20}$ cm$^{-2}$. The FWHM is indicated by the circle,
           $30.5^{\prime\prime} \times 27.5 ^{\prime\prime} $.  }
\vspace{3cm}
         \label{Fig}
   \end{figure}
%

   \begin{figure}
   \centering
   \includegraphics[width=7cm]{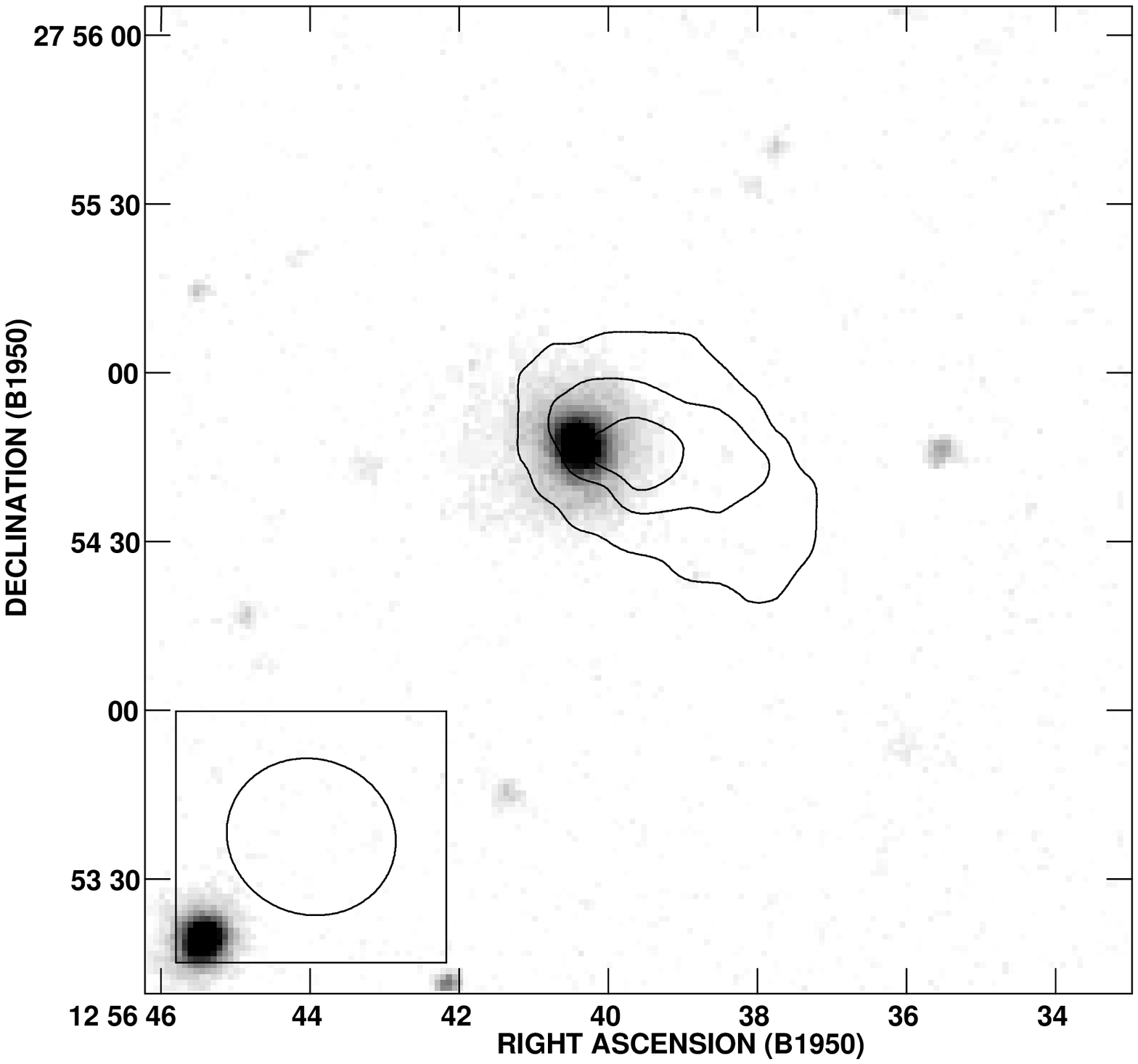}
      \caption{ H{\sc i} density distribution of
           Mrk~058, superposed on a DSS B-band gray scale image. The contours
           are 0.2 (2.5 $\sigma$), 0.7, and $1.1 \times 10^{20}$
           cm$^{-2}$. The FWHM is indicated by the circle,
           $30.3^{\prime\prime} \times 27.6 ^{\prime\prime} $.  } \label{Fig}
           \end{figure}
%

\newpage

   \begin{figure}
   \centering
   \includegraphics[width=7cm]{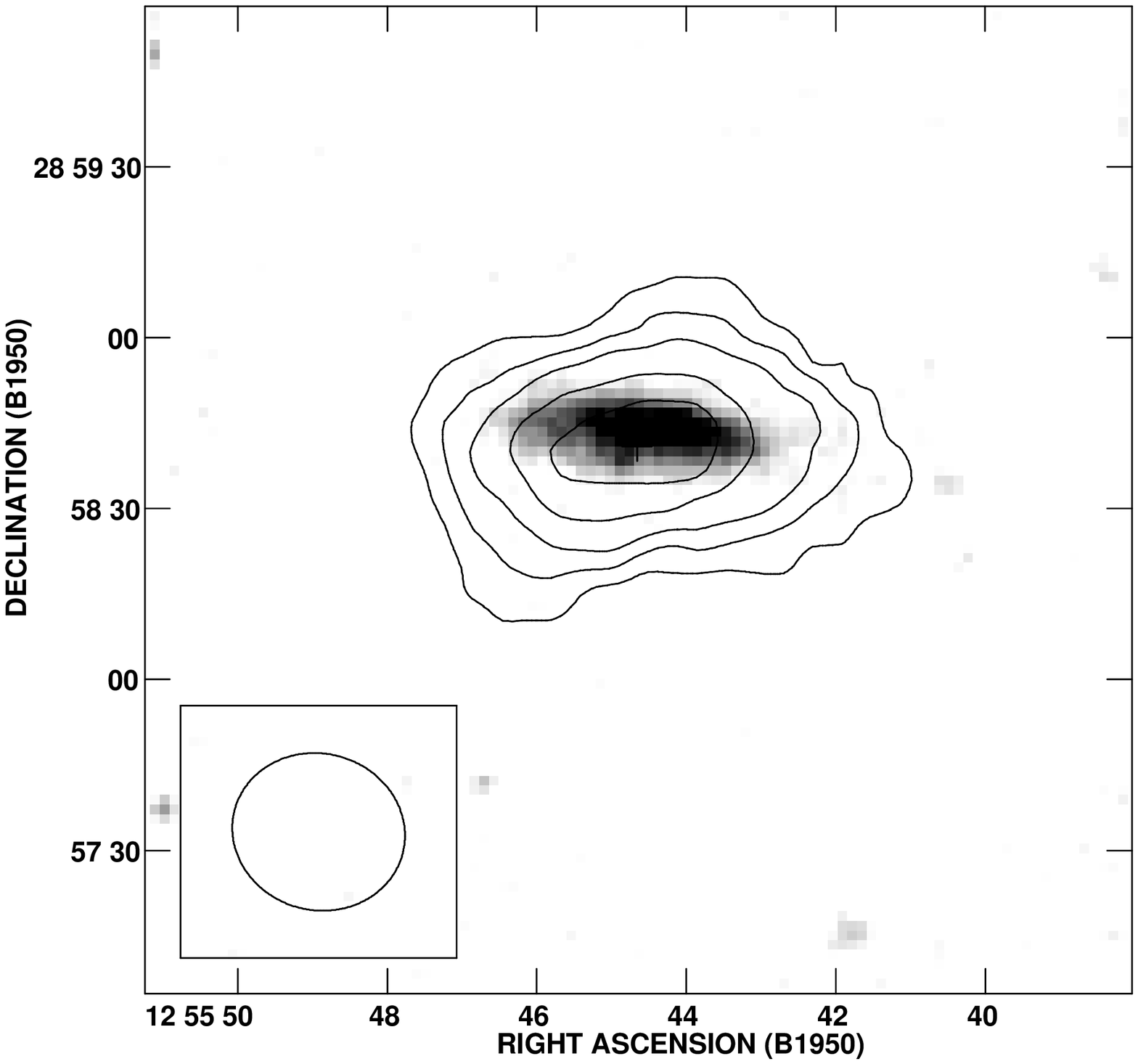}
      \caption{ H{\sc i} density distribution of
        CGCG~160--058, superposed on a DSS B-band gray scale image. The
        contours are 0.4 (2.5 $\sigma$), 1.6, 3.2, 6.4, and $9.6 \times
        10^{20}$ cm$^{-2}$. The FWHM is indicated by the circle,
        $30.5^{\prime\prime} \times 27.5 ^{\prime\prime} $.  } \vspace{3cm}
        \label{Fig} \end{figure}
%

   \begin{figure}
   \centering
   \includegraphics[width=7cm]{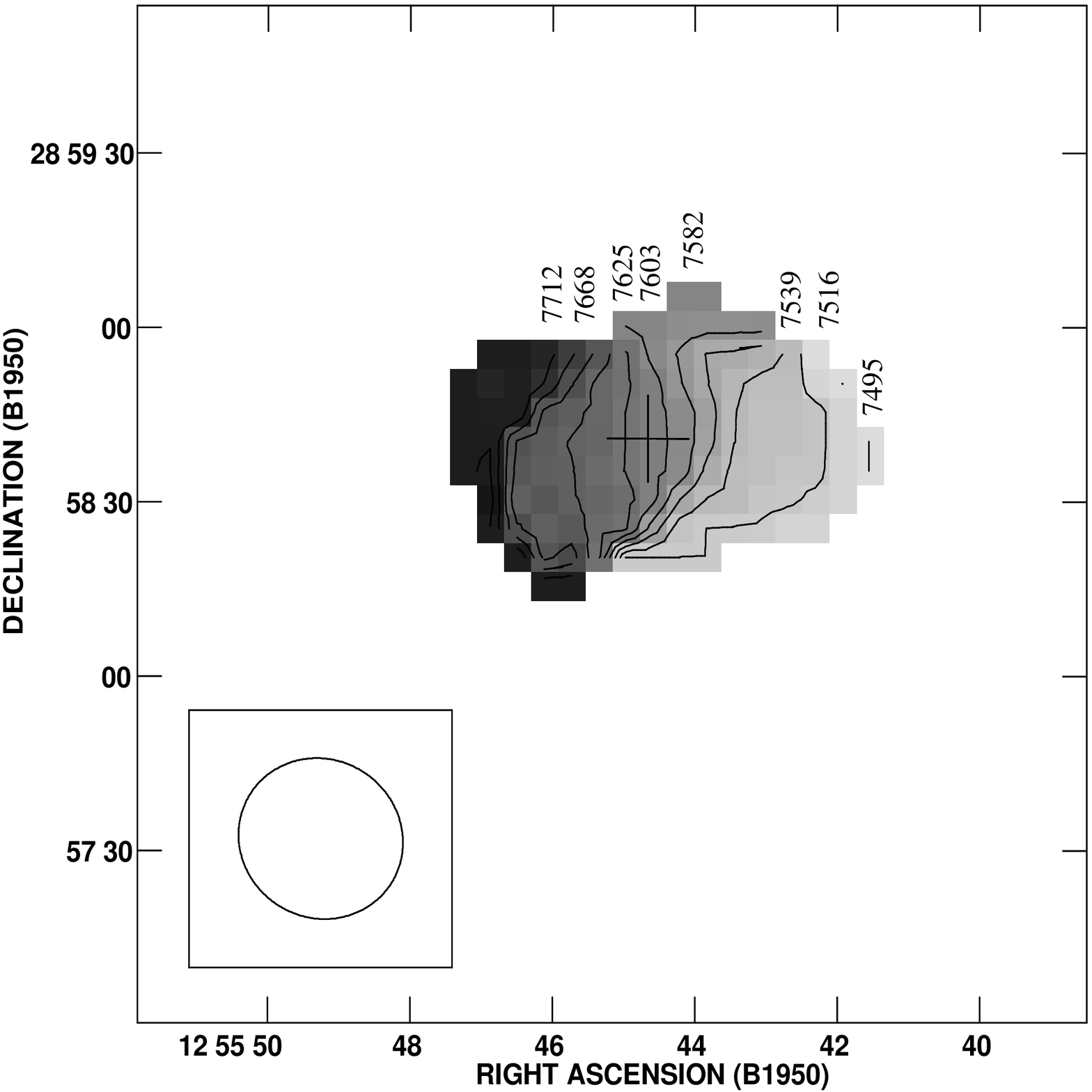}
      \caption{ Intensity weighted mean velocity
           field of CGCG~160--058.  The optical center of the galaxy is
           indicated with a cross.  The numbers indicate heliocentric
           velocity in \km. The FWHM is indicated by the circle,
           $30.5^ {\prime\prime} \times 27.5^{\prime\prime}$.  } \label{Fig}
           \end{figure}
%

\newpage

   \begin{figure}
   \centering
   \includegraphics[width=7cm]{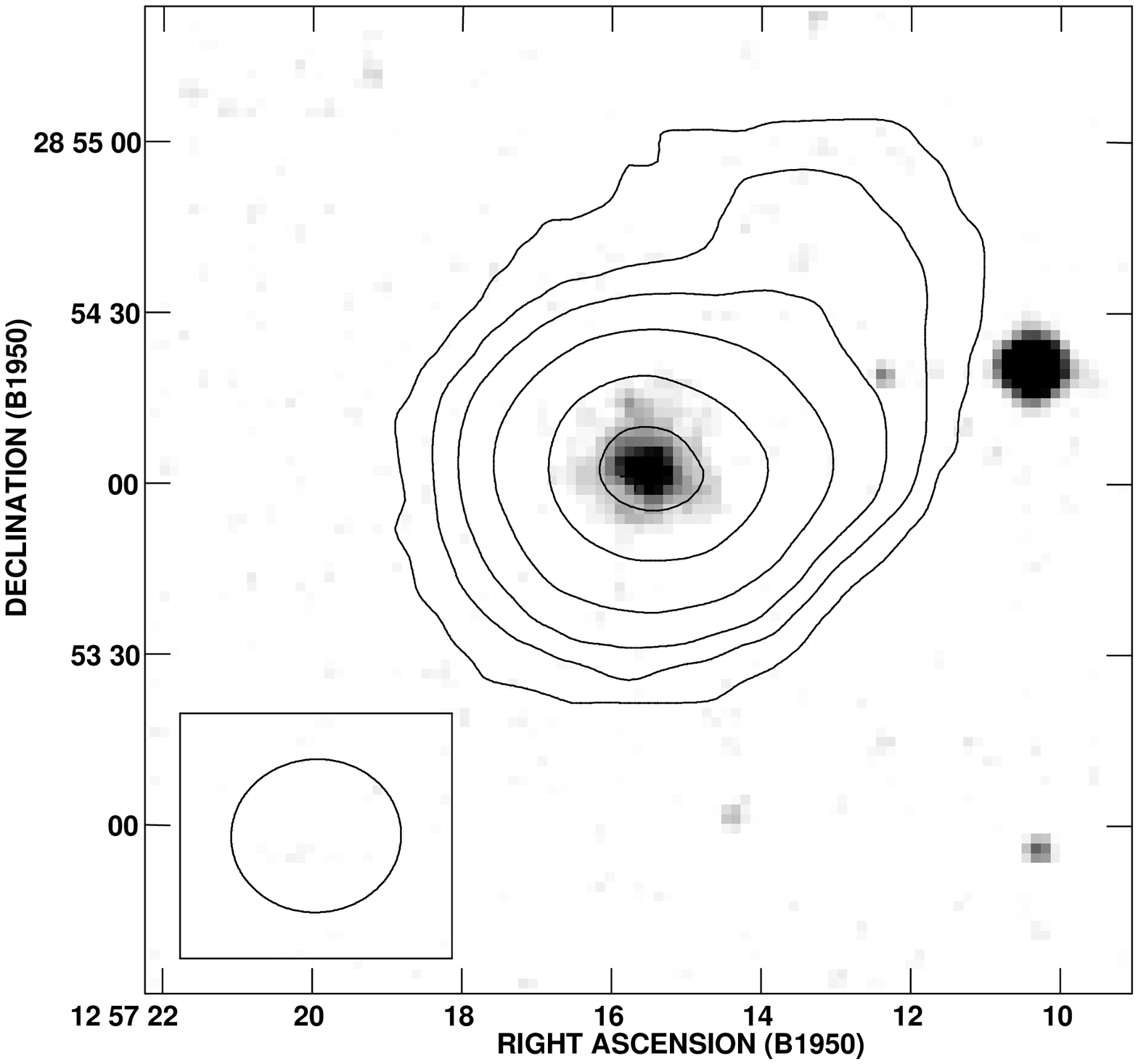}
      \caption{ H{\sc i} density distribution of
                CGCG~160-076, 
           superposed on a DSS B-band gray scale image. The contours are 0.2
           (2.5 $\sigma$), 0.8, 1.6, 3.3, 6.6 and $9.9 \times 10^{20}$
           cm$^{-2}$. The FWHM is indicated by the  circle,
           $29.9^{\prime\prime} \times 26.9^{\prime\prime} $.
              }
    \vspace{3cm}
       \label{Fig}
   \end{figure}
%

   \begin{figure}
   \centering
   \includegraphics[width=7cm]{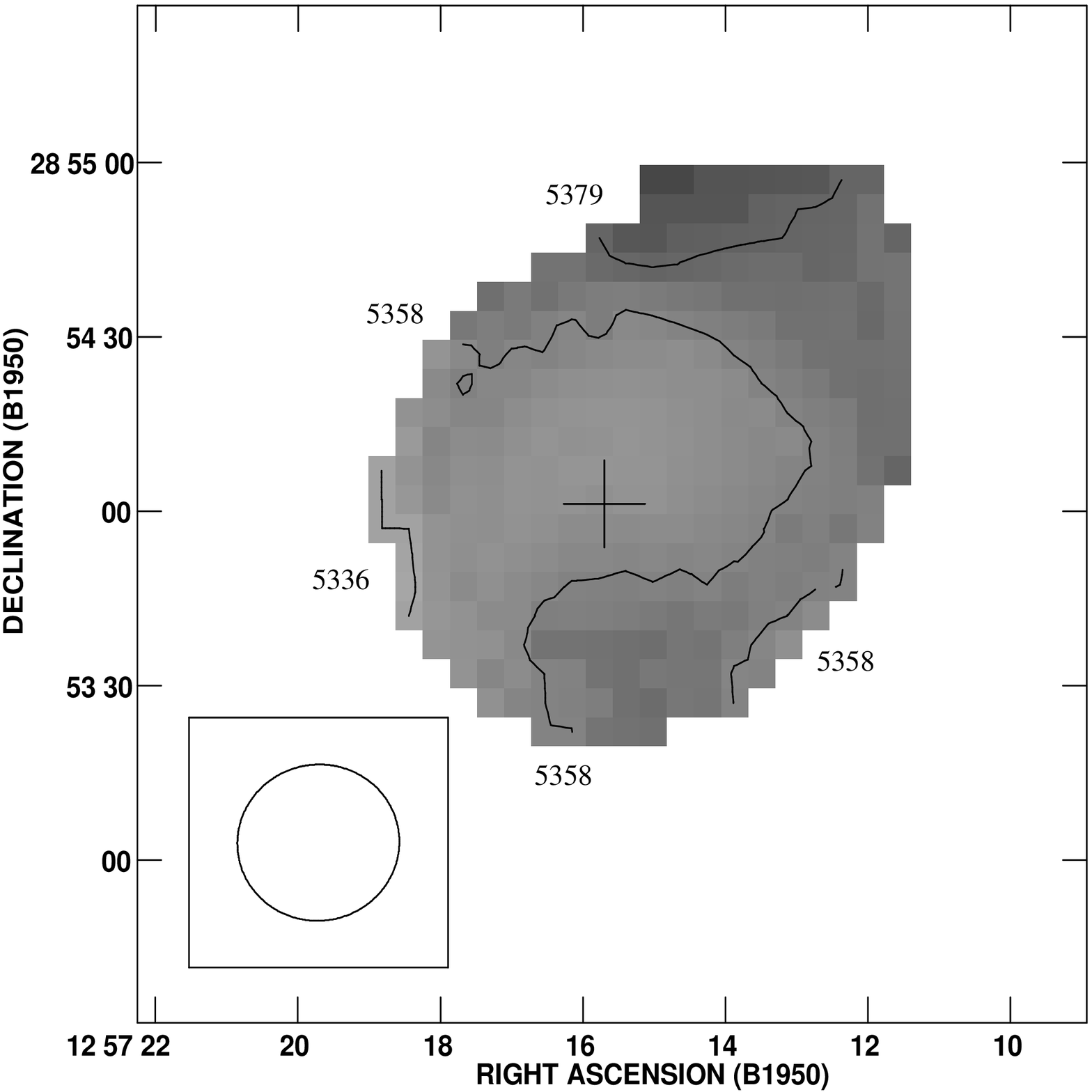}
      \caption{ Intensity weighted mean velocity field
           of CGCG~160-076.  The optical center of the galaxy is indicated
           with a cross.  The numbers indicate heliocentric velocity in \km.
           The FWHM is indicated by the circle, $29.9^
           {\prime\prime} \times 26.9^{\prime\prime}$.
              }
         \label{Fig}
   \end{figure}
%

\newpage

   \begin{figure}
   \centering
   \includegraphics[width=7cm]{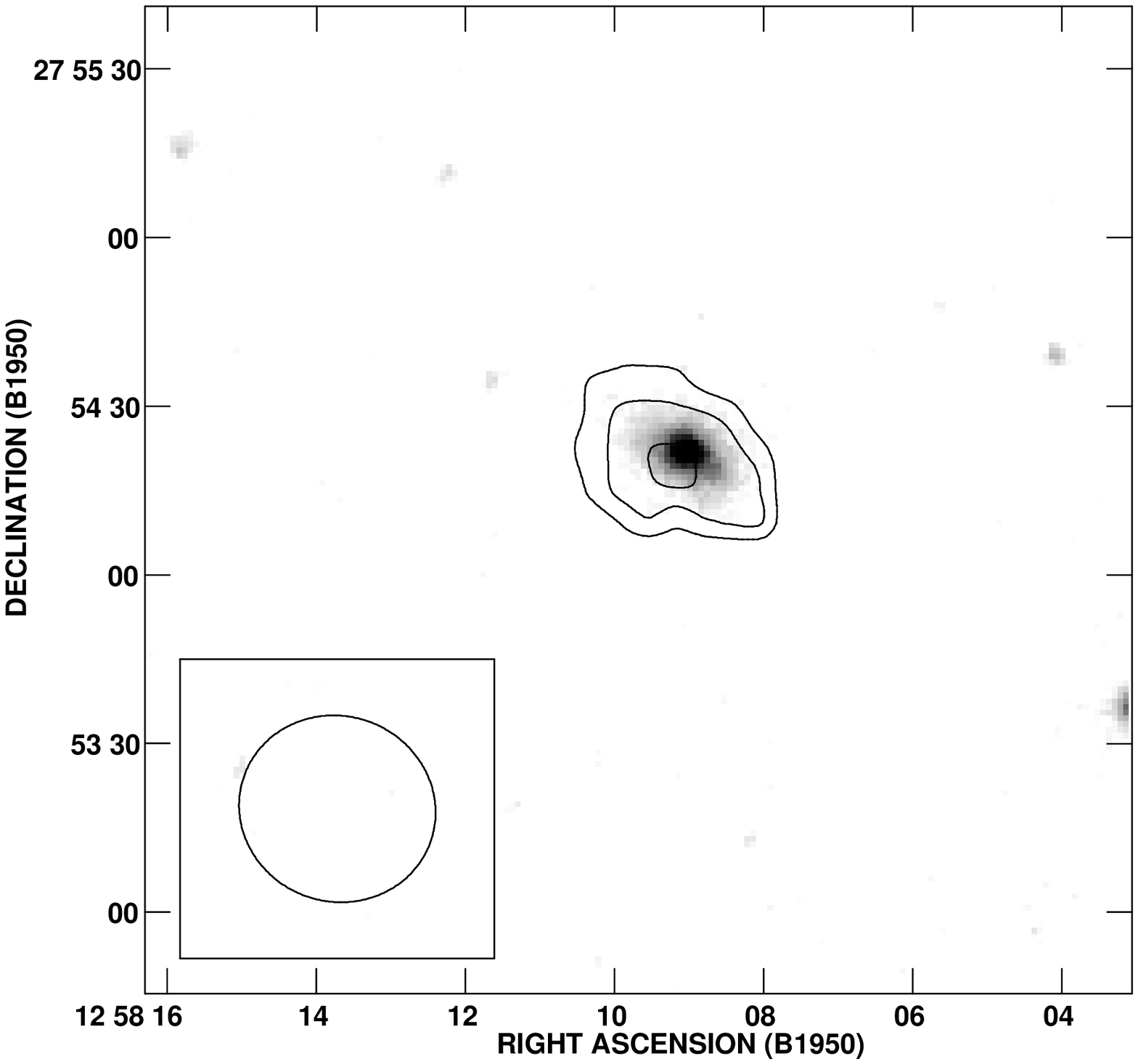}
      \caption{ H{\sc i} density distribution of
                CGCG~160-086, 
           superposed on a DSS B-band gray scale image. The contours are 0.4
           (2.5 $\sigma$), 0.9, and $1.3 \times 10^{20}$ cm$^{-2}$. The FWHM
           is indicated by the  circle, $35.2^{\prime\prime} \times
           33.0 ^{\prime\prime} $.
              }
\vspace{15cm}
         \label{Fig}
   \end{figure}
%

\newpage

   \begin{figure}
   \centering
   \includegraphics[width=7cm]{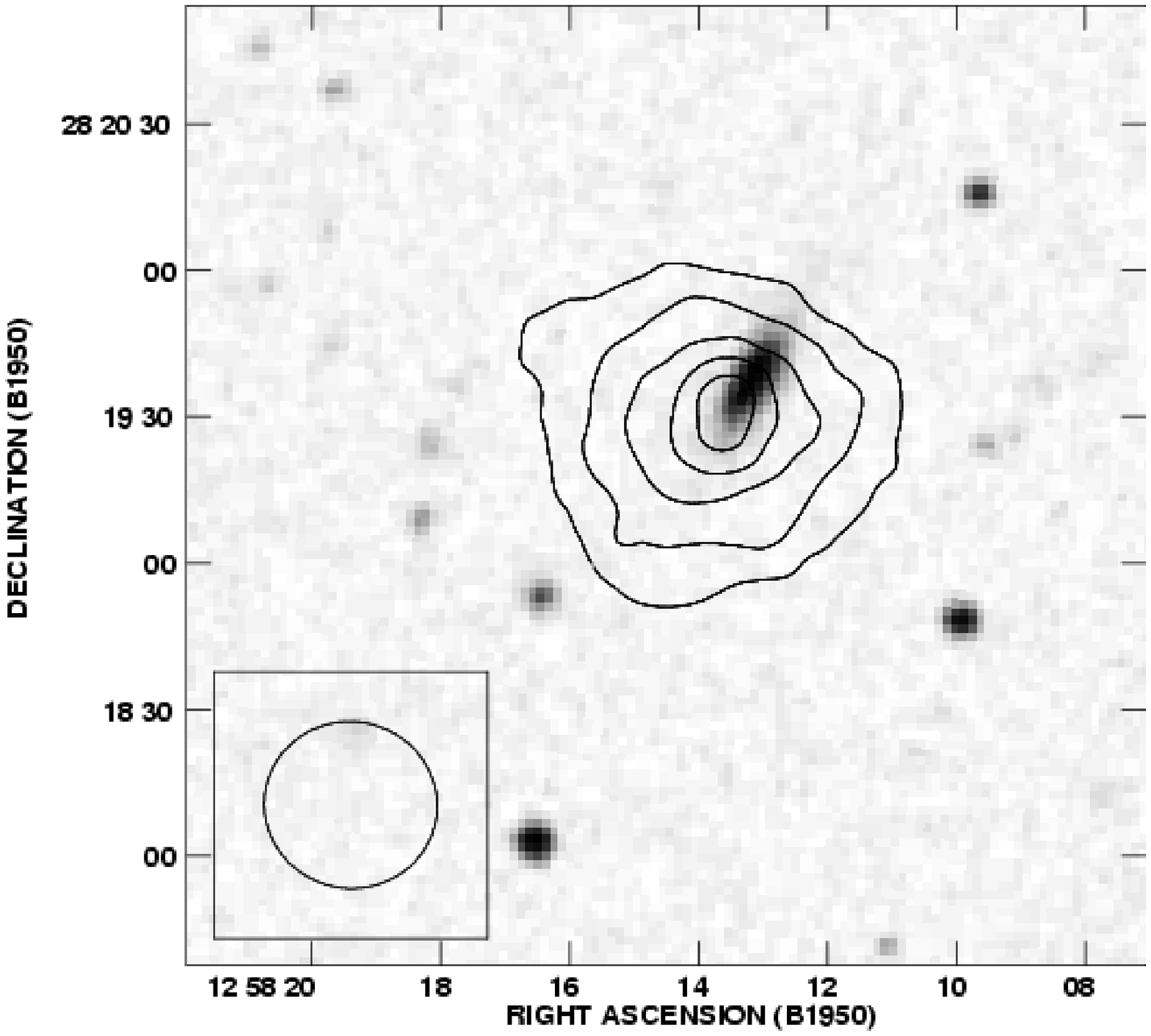}
      \caption{ H{\sc
        i} density distribution of IC~4040, superposed on a DSS B-band gray
        scale image.  The contours are 0.2 (5$\sigma$), 0.4, 0.7, 1.1, and
        $1.4 \times 10^{20}$ cm$^{-2}$.  The FWHM is indicated by the circle,
        $35.5^{\prime\prime} \times 34.1 ^{\prime\prime} $.
              }
    \vspace{3cm}
       \label{Fig}
   \end{figure}
%

   \begin{figure}
   \centering
   \includegraphics[width=7cm]{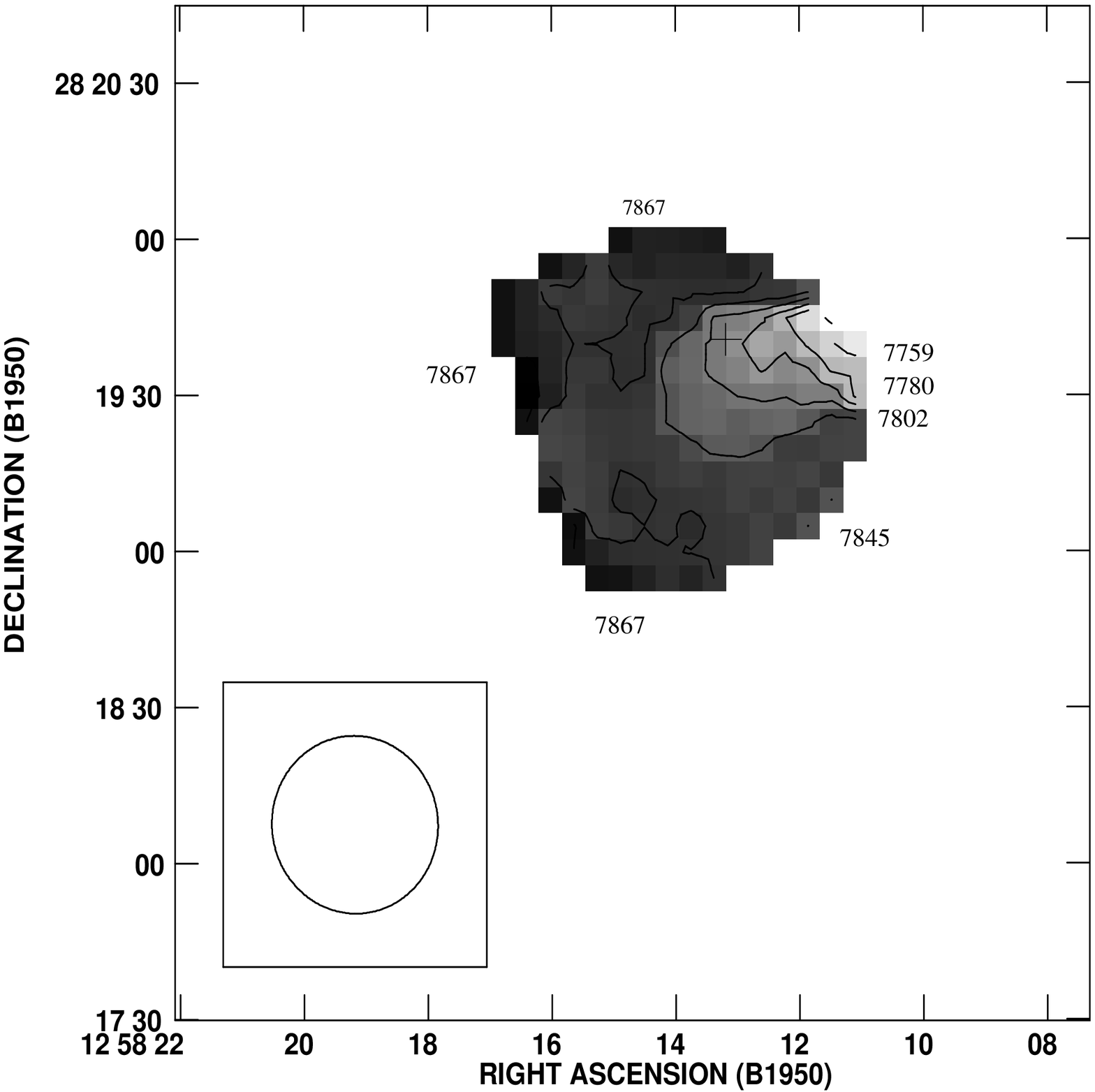}
      \caption{ Intensity weighted mean velocity field
           of IC~4040.  The optical center of the galaxy is indicated
           with a cross. The numbers indicate heliocentric velocity in \km.
           The FWHM is indicated by the circle, $35.5^
           {\prime\prime} \times 34.1^{\prime\prime}$.
              }
         \label{Fig}
   \end{figure}
%

\newpage

   \begin{figure}
   \centering
   \includegraphics[width=7cm]{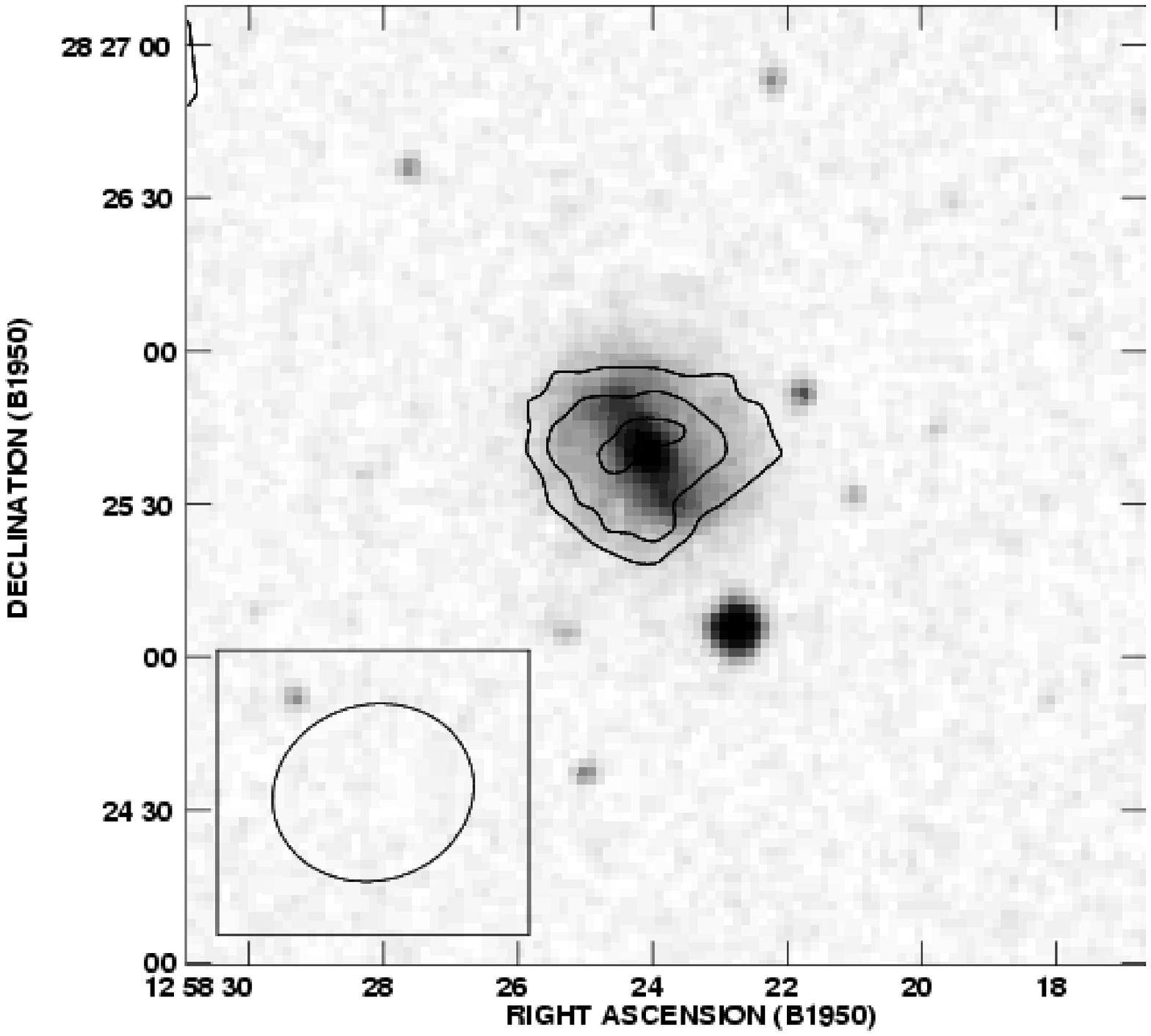}
      \caption{ H{\sc i} density distribution of
           NGC~4907, superposed on a DSS B-band gray scale image. The
           contours are 0.2 (2.5 $\sigma$), 0.5, and $0.9 \times 10^{20}$
           cm$^{-2}$. The FWHM is indicated by the circle,
           $39.8^{\prime\prime} \times 34.5 ^{\prime\prime} $.}
\vspace{15cm}
         \label{Fig}
   \end{figure}
%

\newpage

   \begin{figure}
   \centering
   \includegraphics[width=7cm]{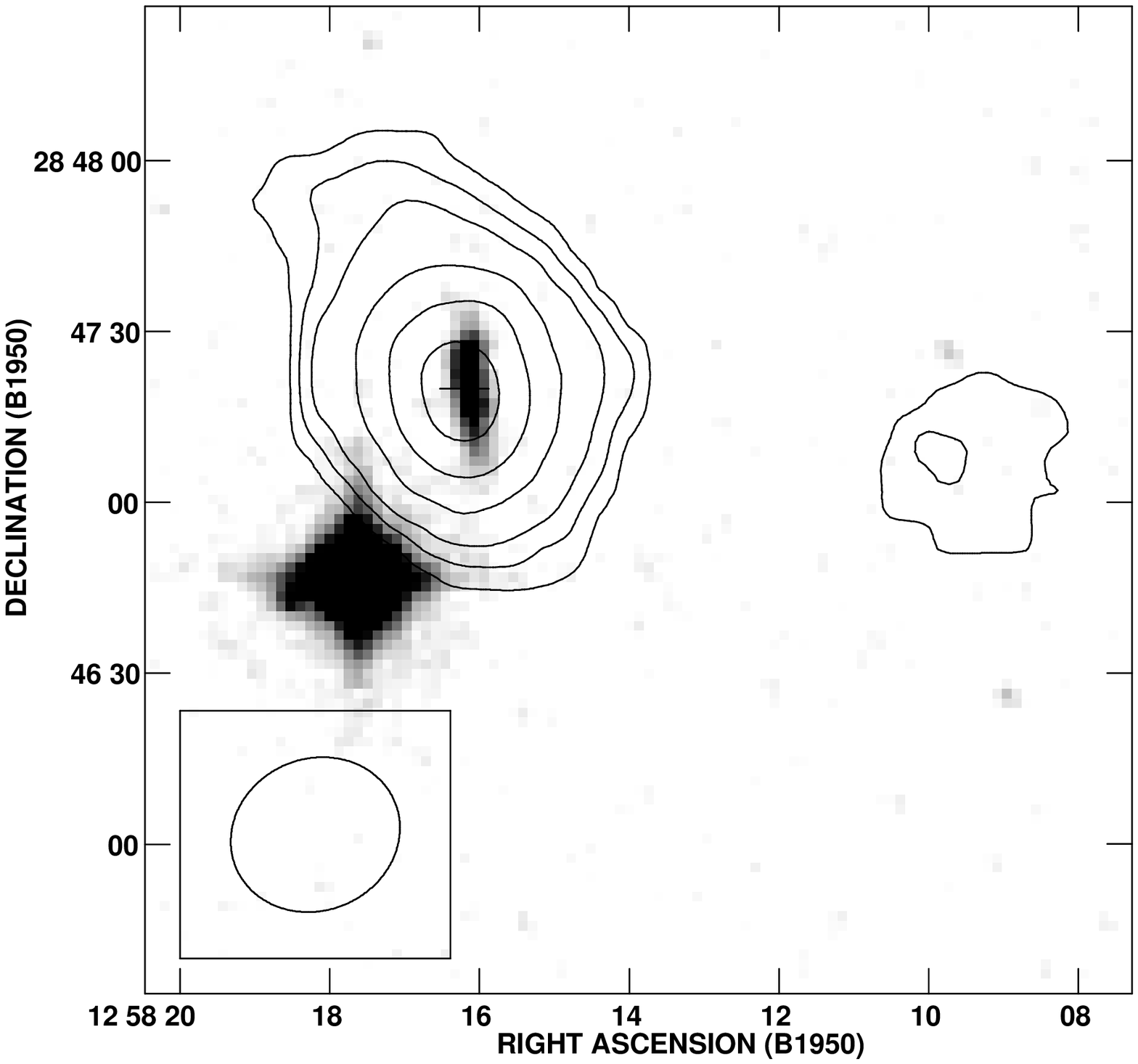}
      \caption{ H{\sc
        i} density distribution of KUG~1258+287, superposed on a DSS B-band
        gray scale image. The contours are 0.3 (2.5 $\sigma$), 1.1, 2.2, 4.5,
        6.8, and $9.0 \times 10^{20}$ cm$^{-2}$. The FWHM is indicated by the
        circle, $30.2^{\prime\prime} \times 26.7 ^{\prime\prime} $.
              }
    \vspace{3cm}
       \label{Fig}
   \end{figure}
%

   \begin{figure} \centering \includegraphics[width=7cm]{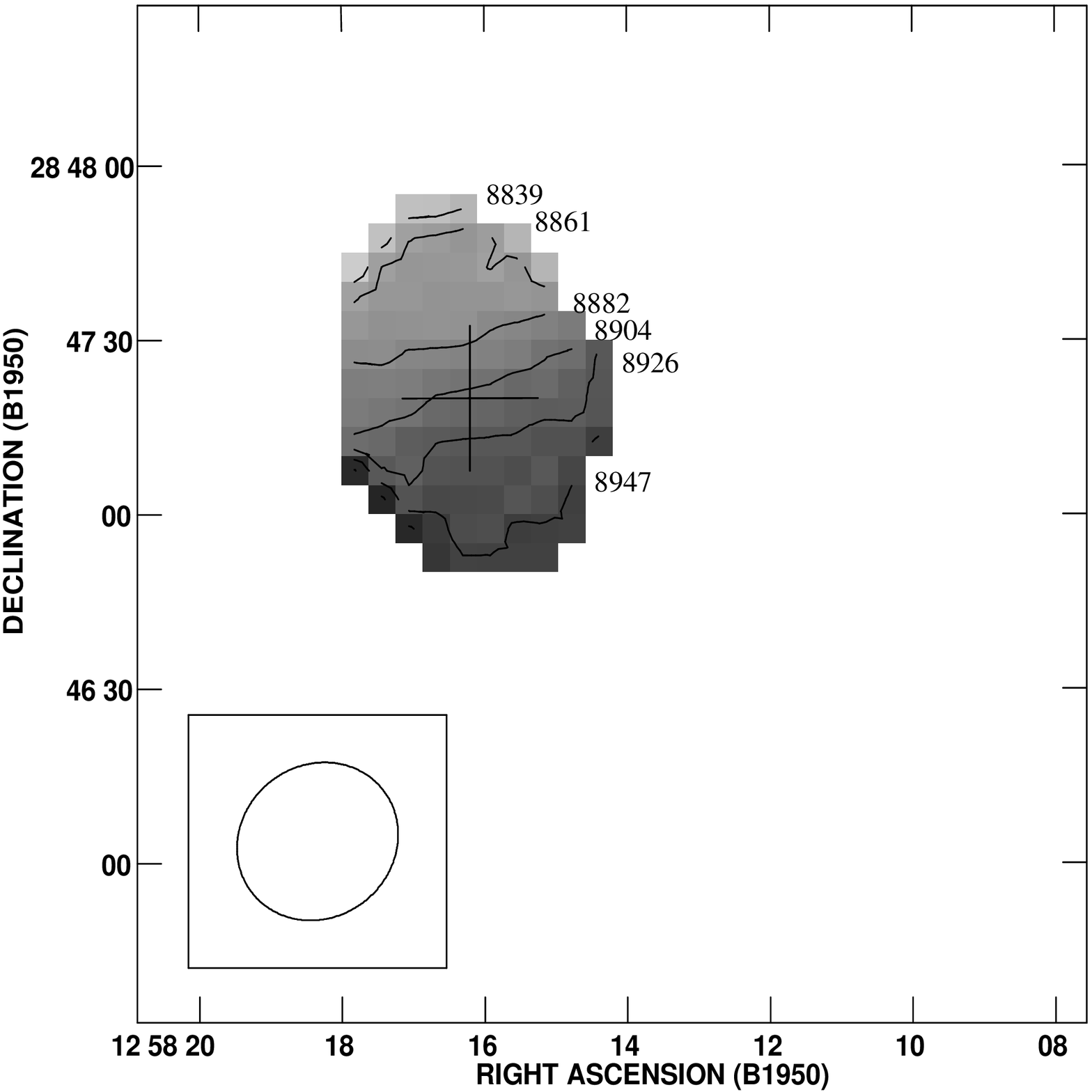}
   \caption{ Intensity weighted mean velocity field
   of KUG~1258+287.  The optical center of the galaxy is indicated with a
   cross.  The numbers indicate heliocentric velocity in \km.  The FWHM is
   indicated by the circle, $30.2^ {\prime\prime} \times
   26.7^{\prime\prime}$.  } \label{Fig} \end{figure}
%

   \begin{figure}
   \centering
   \includegraphics[width=7cm]{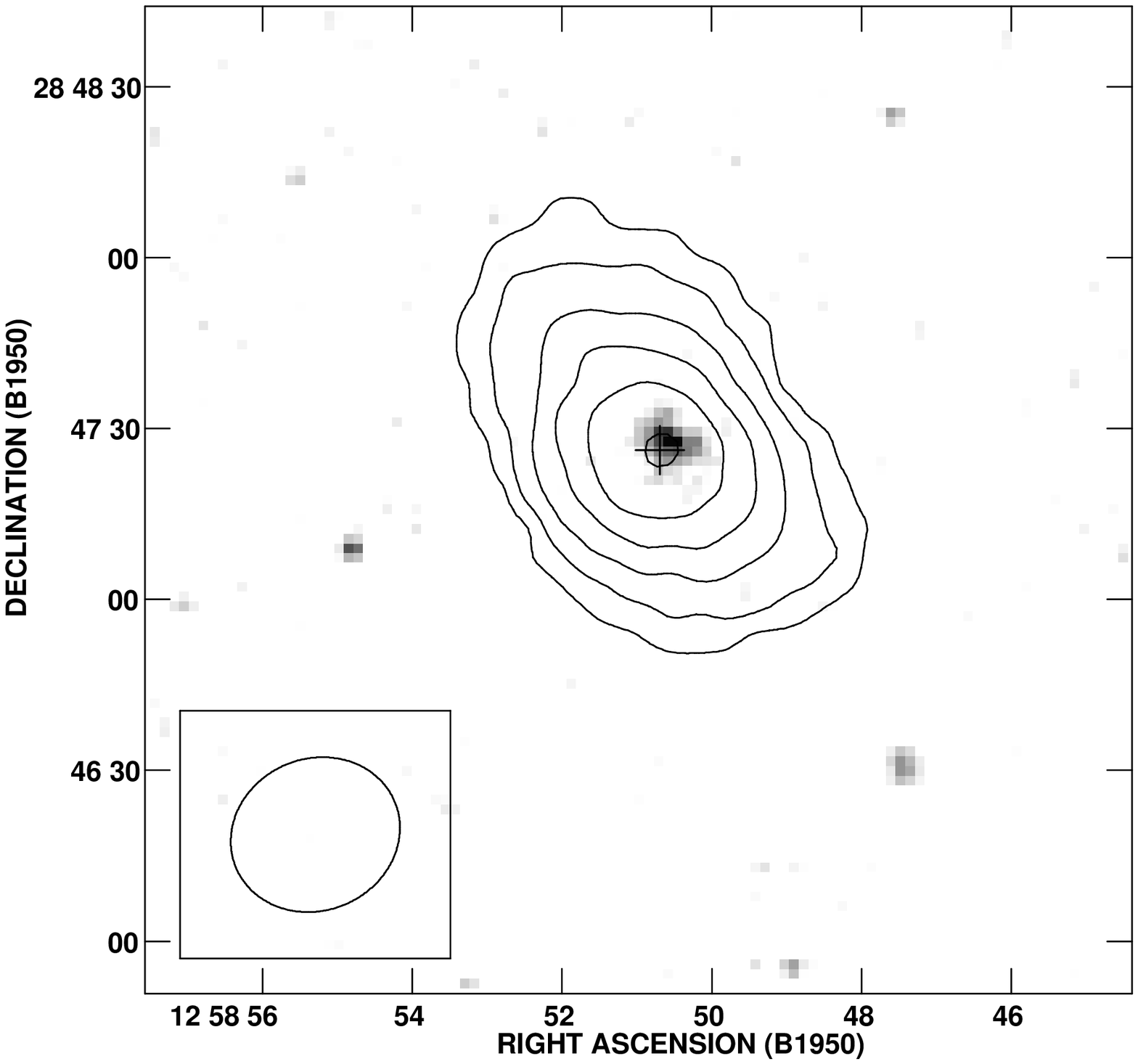}
      \caption{ H{\sc i} density distribution of
        FOCA~0195, superposed on a DSS B-band gray scale image. The contours
        are 0.2 (2.5 $\sigma$), 0.9, 1.8, 2.6, 3.5, and $4.4 \times 10^{20}$
        cm$^{-2}$. The FWHM is indicated by the circle, $30.2^{\prime\prime}
        \times 26.7 ^{\prime\prime} $.
              }
    \vspace{3cm}
       \label{Fig}
   \end{figure}
%

   \begin{figure} \centering \includegraphics[width=7cm]{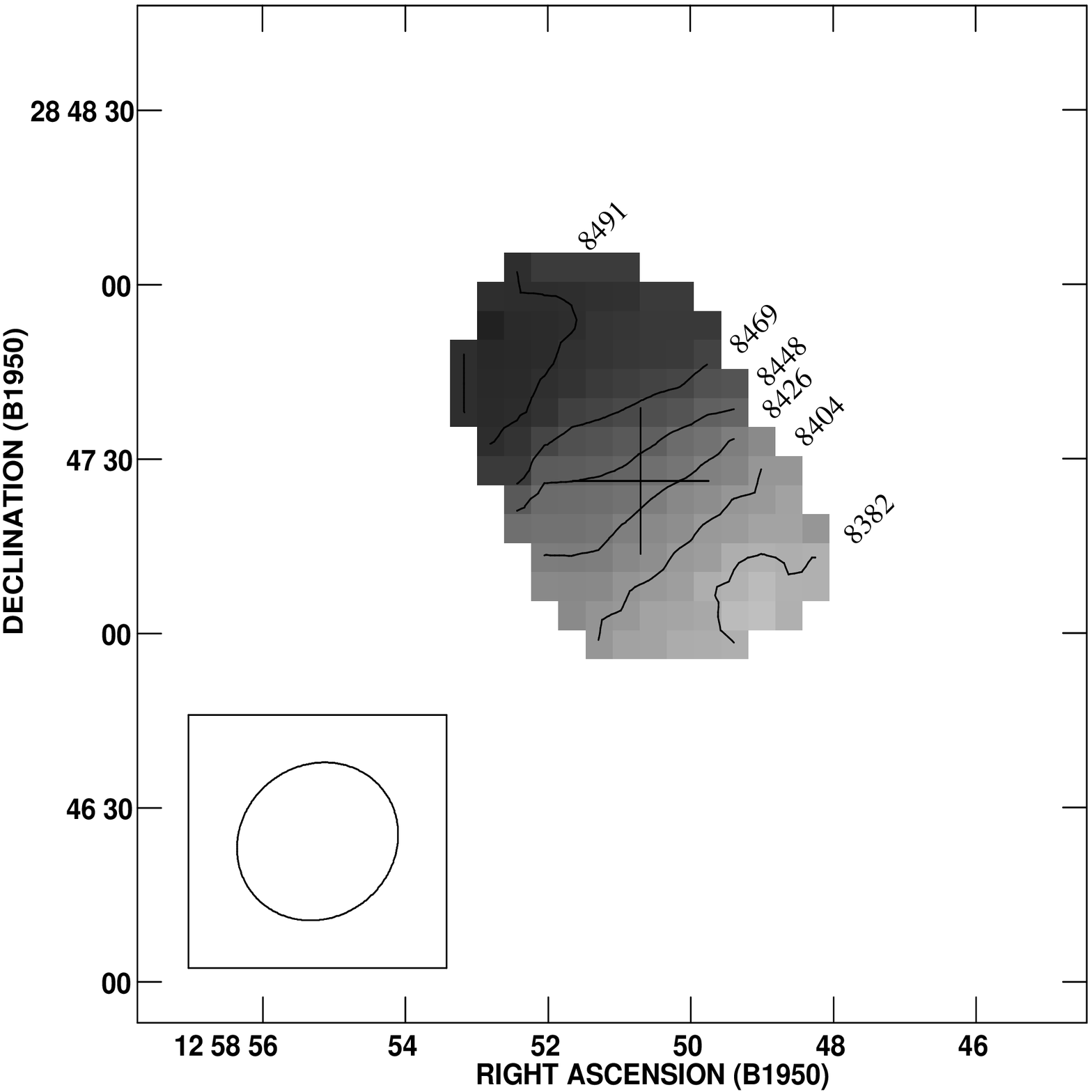}
   \caption{ Intensity weighted mean velocity field
   of FOCA~0195.  The optical center of the galaxy is indicated with a cross.
   The numbers indicate heliocentric velocity in \km.  The FWHM is indicated
   by the circle, $30.2^ {\prime\prime} \times 26.7^{\prime\prime}$.  }
   \label{Fig} \end{figure}
%

\newpage

   \begin{figure}
   \centering
   \includegraphics[width=7cm]{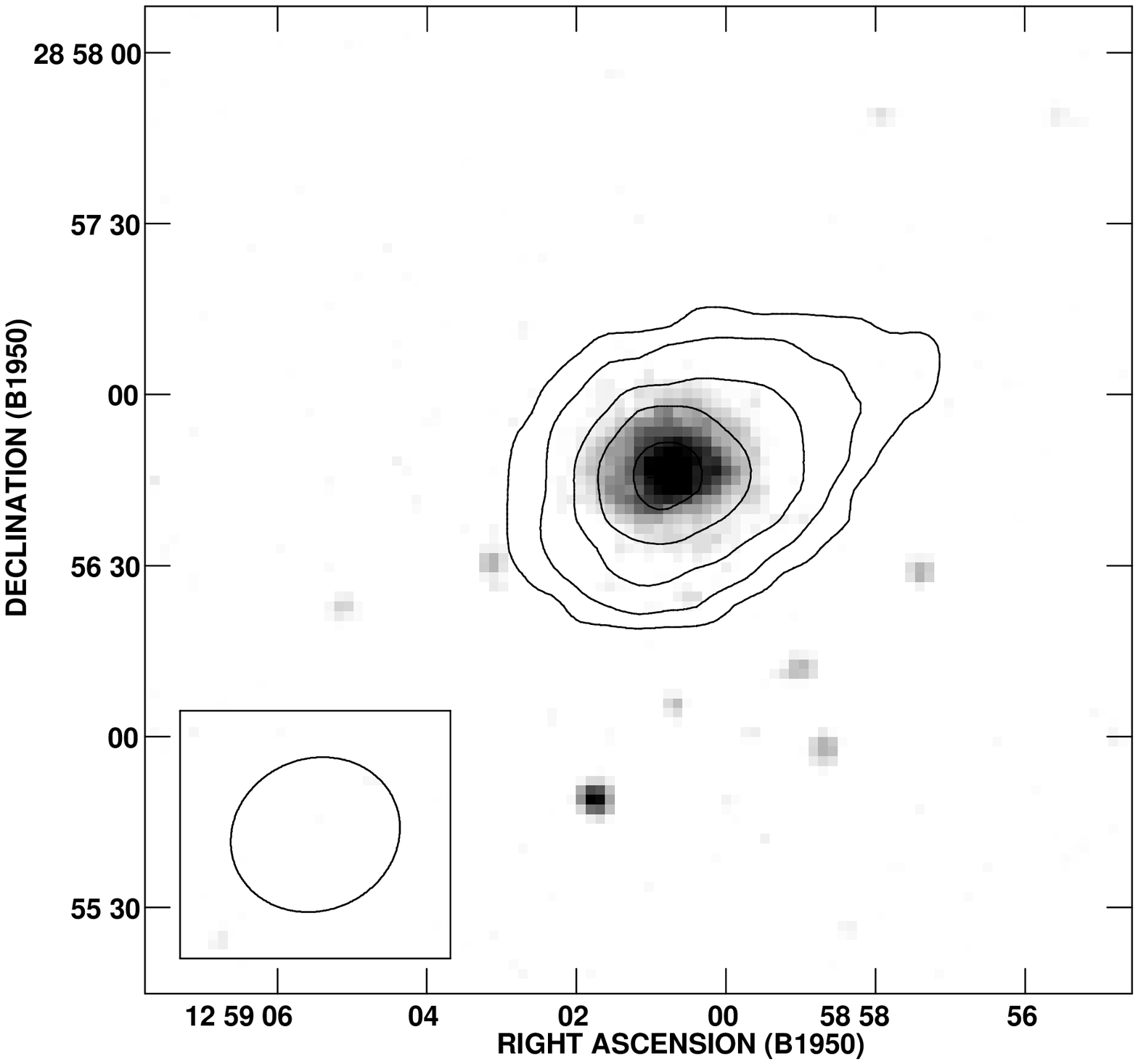}
      \caption{ H{\sc i} density distribution of
        CGCG~160-098, superposed on a DSS B-band gray scale image. The
        contours are 0.3 (2.5 $\sigma$), 1.1, 2.2, and $4.4 \times 10^{20}$
        cm$^{-2}$. The FWHM is indicated by the circle, $30.2^{\prime\prime}
        \times 26.7 ^{\prime\prime} $.
              }
    \vspace{3cm}
       \label{Fig}
   \end{figure}
%

   \begin{figure} \centering \includegraphics[width=7cm]{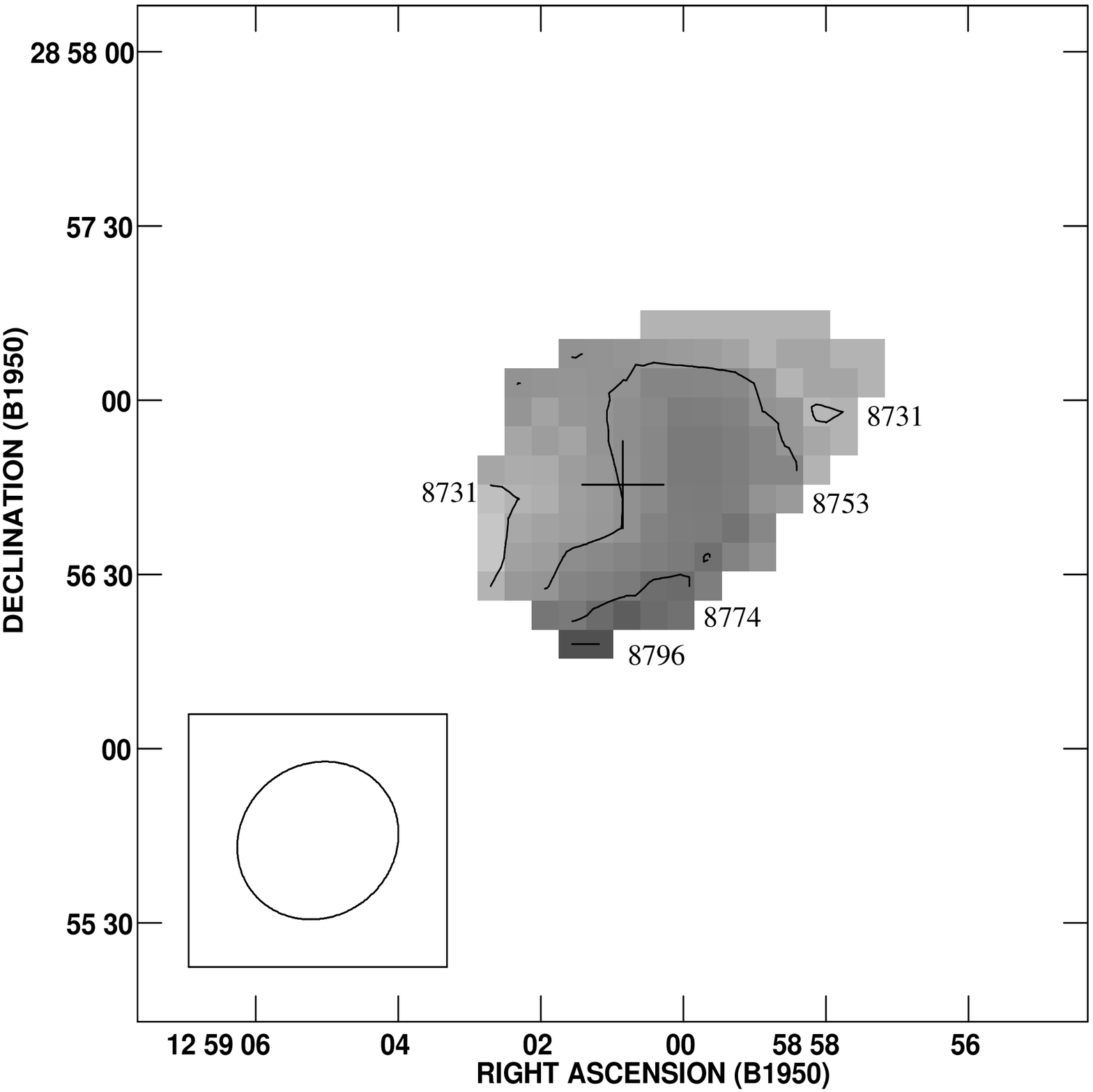}
   \caption{ Intensity weighted mean velocity field
   of CGCG~160-098.  The optical center of the galaxy is indicated with a
   cross.  The numbers indicate heliocentric velocity in \km.  The FWHM is
   indicated by the circle, $30.2^ {\prime\prime} \times
   26.7^{\prime\prime}$.  } \label{Fig} \end{figure}
%

\newpage

   \begin{figure}
   \centering
   \includegraphics[width=7cm]{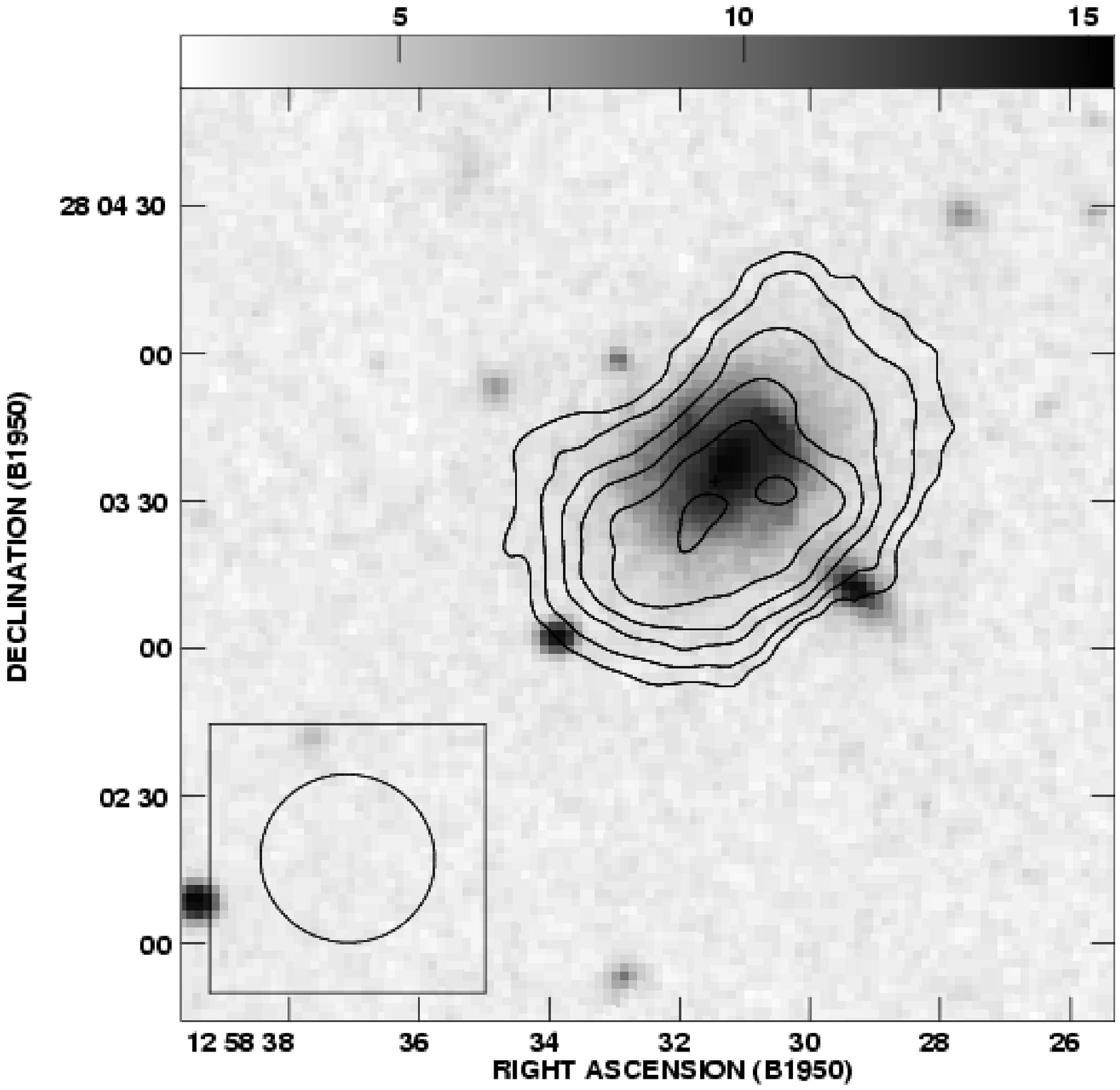}
      \caption{ H{\sc i} density distribution of
        NGC~4911, superposed on a DSS B-band gray scale image. The contours
        are 0.3 (2.5 $\sigma$), 1.4, and $2.0 \times 10^{20}$ cm$^{-2}$. The
        FWHM is indicated by the circle, $29.8^{\prime\prime} \times
        27.7^{\prime\prime} $.
              }
    \vspace{3cm}
       \label{Fig}
   \end{figure}
%

   \begin{figure} \centering \includegraphics[width=7cm]{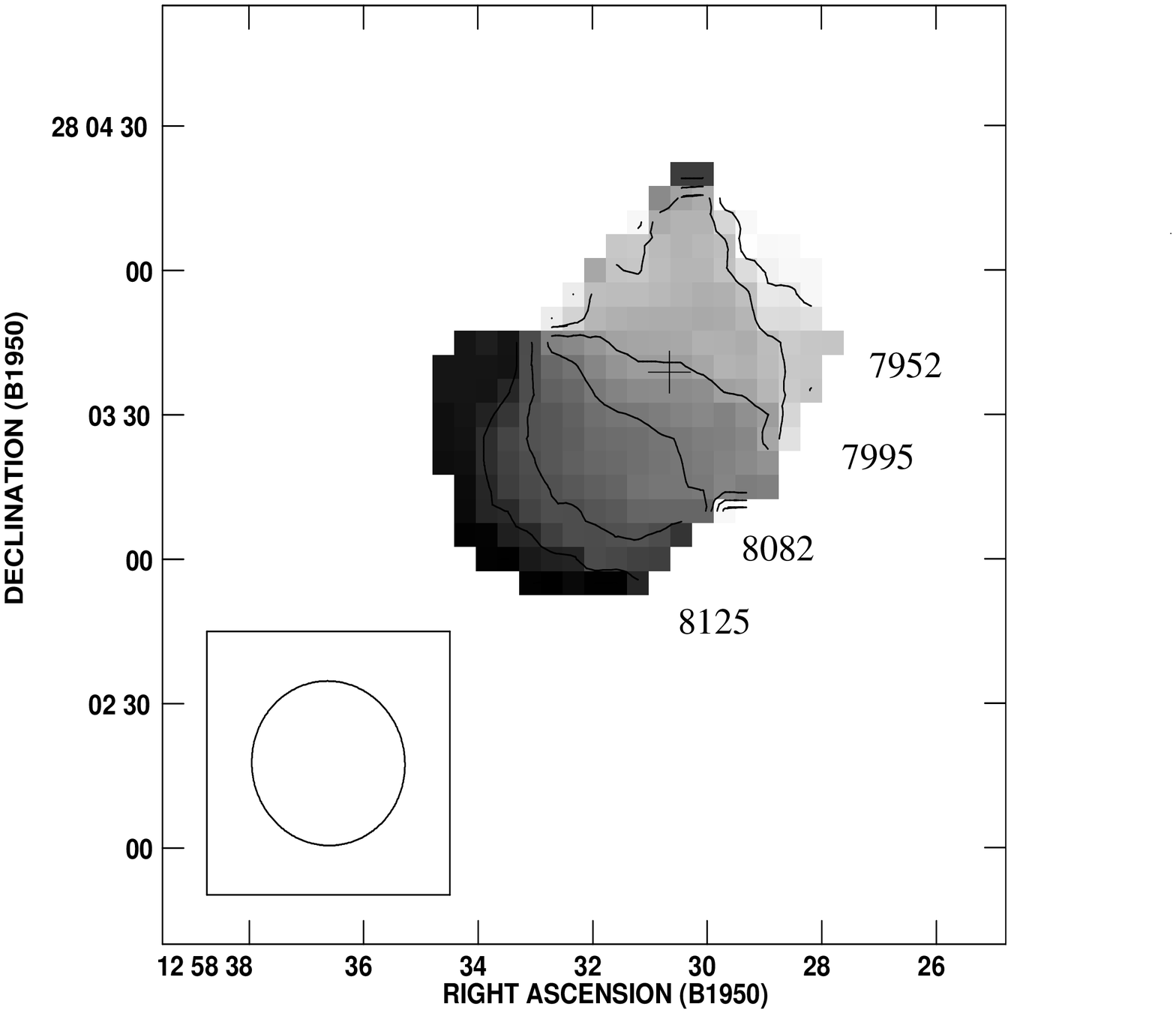}
   \caption{ Intensity weighted mean velocity field
   of NGC~4911. The optical center of the galaxy is indicated with a
   cross. The numbers indicate heliocentric velocity in \km.  The FWHM is
   indicated by the circle, $29.8^ {\prime\prime} \times
   27.7^{\prime\prime}$.  } \label{Fig} \end{figure}
%

\newpage

   \begin{figure}
   \centering
   \includegraphics[width=7cm]{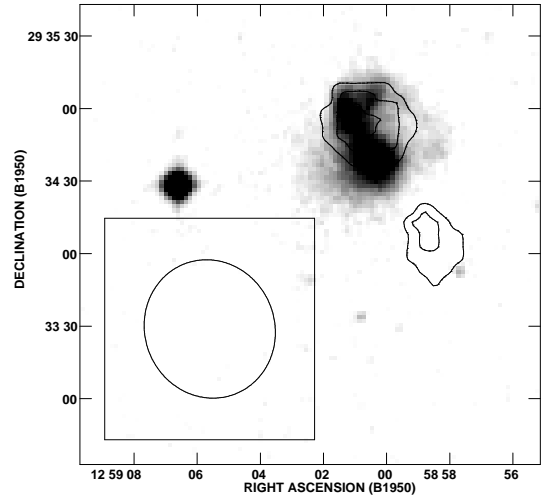}
      \caption{ H{\sc i} density distribution showing
        the emission feature of NGC~4922, superposed on a DSS B-band gray
        scale image. Strong \hi~absorption is present which attenuates the 
	emission.  The contours are
        0.2 (2.5 $\sigma$), 0.4, and $0.5 \times 10^{20}$ cm$^{-2}$. The FWHM
        is indicated by the circle, $57.8^{\prime\prime} \times
        53.4^{\prime\prime} $.
              }
\vspace{15cm}
         \label{Fig}
   \end{figure}
%

\newpage

   \begin{figure}
   \centering
   \includegraphics[width=7cm]{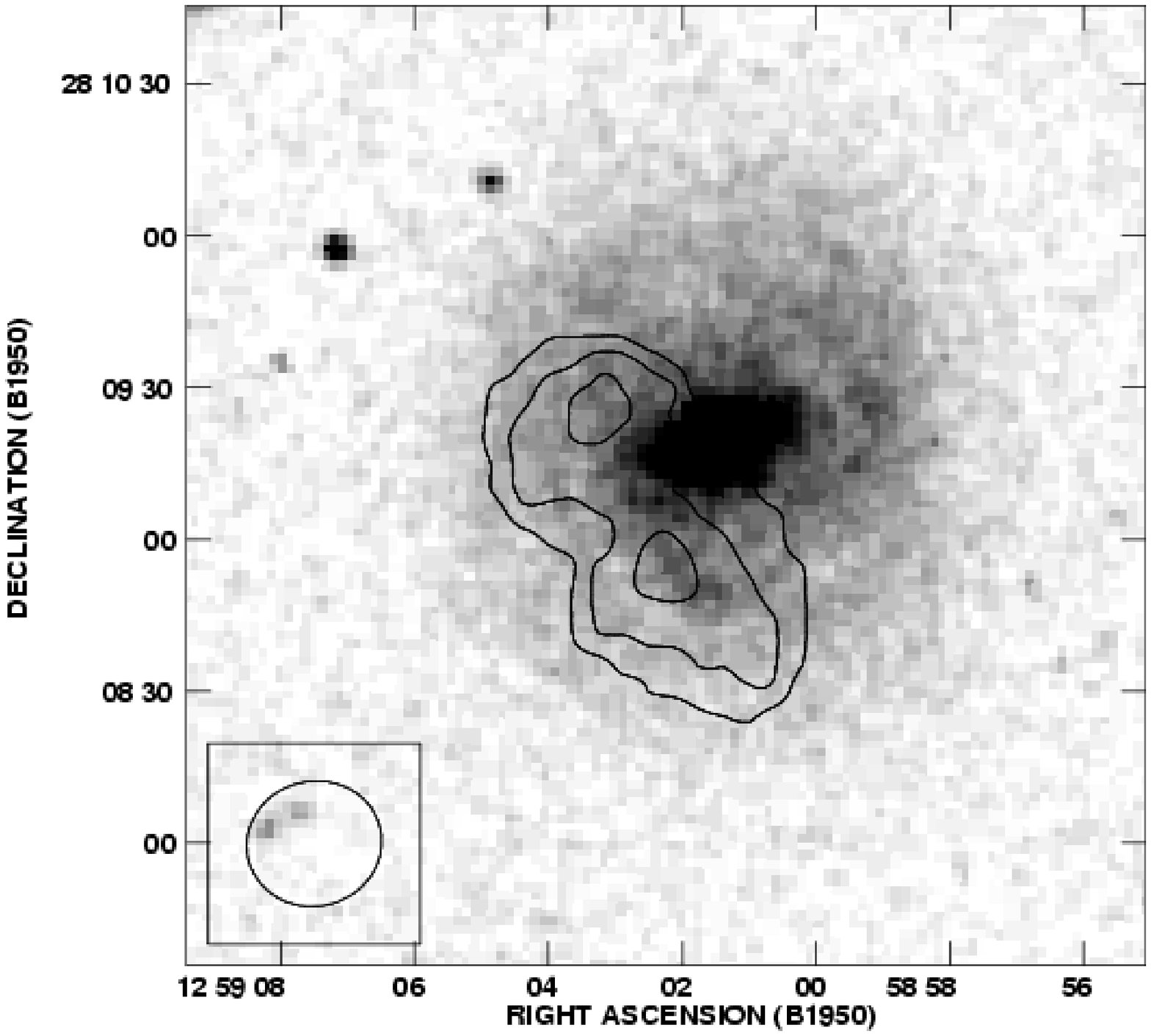}
      \caption{ H{\sc i} density distribution of
        NGC~4921, superposed on a DSS B-band gray scale image. The contours
        are 0.3 (2.5 $\sigma$), 1.2, 1.7, and $2.3 \times 10^{20}$
        cm$^{-2}$. The FWHM is indicated by the circle, $39.8^{\prime\prime}
        \times 34.5 ^{\prime\prime} $.
              }
    \vspace{3cm}
       \label{Fig}
   \end{figure}
%

   \begin{figure} \centering \includegraphics[width=7cm]{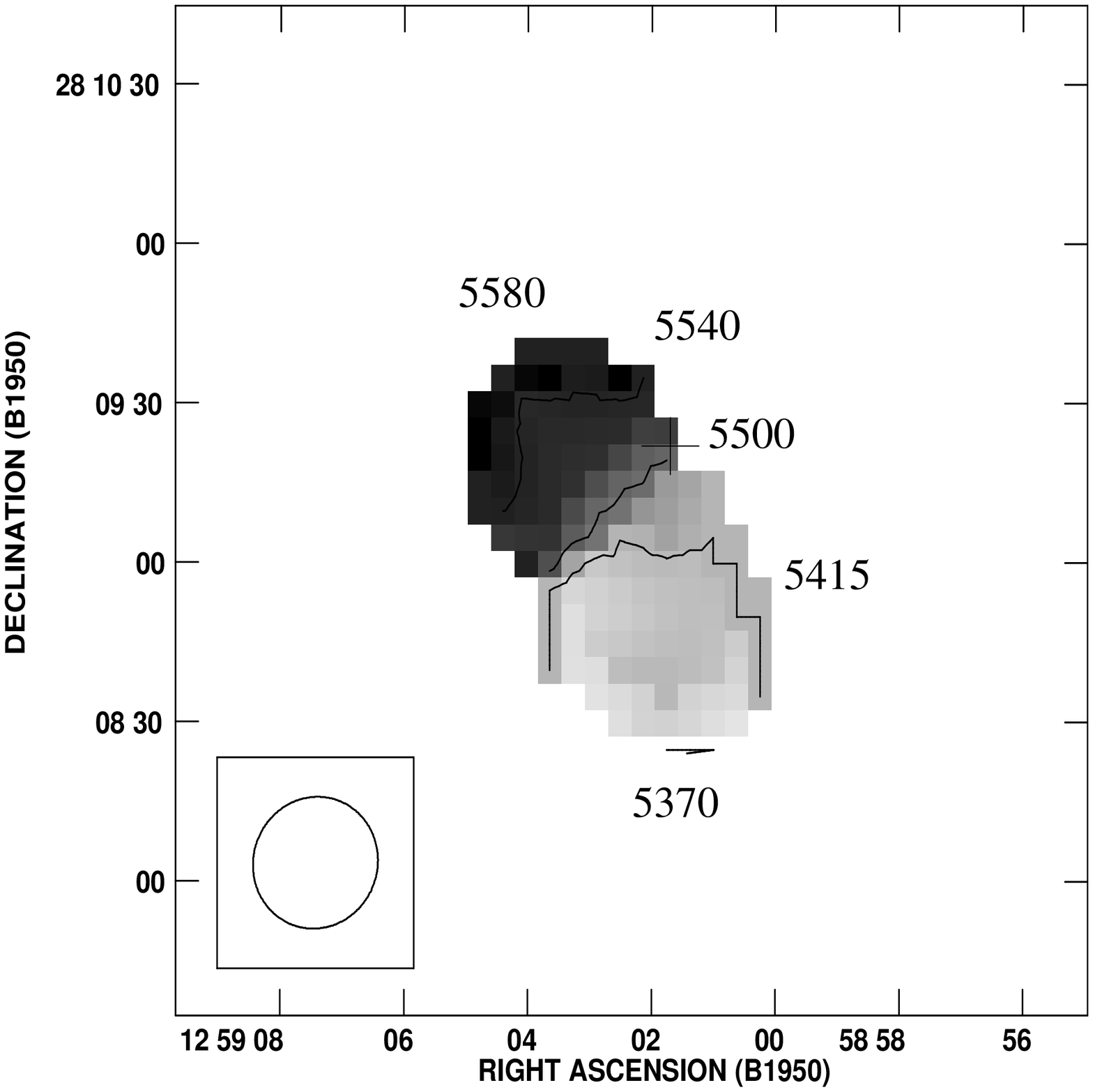}
   \caption{ Intensity weighted mean velocity field
   of NGC~4921. The optical center of the galaxy is indicated with a cross.
   The numbers indicate heliocentric velocity in \km. The FWHM is indicated
   by the circle, $39.8^ {\prime\prime} \times 34.5^{\prime\prime}$.  }
   \label{Fig} \end{figure}
%

\newpage

   \begin{figure}
   \centering
   \includegraphics[width=7cm]{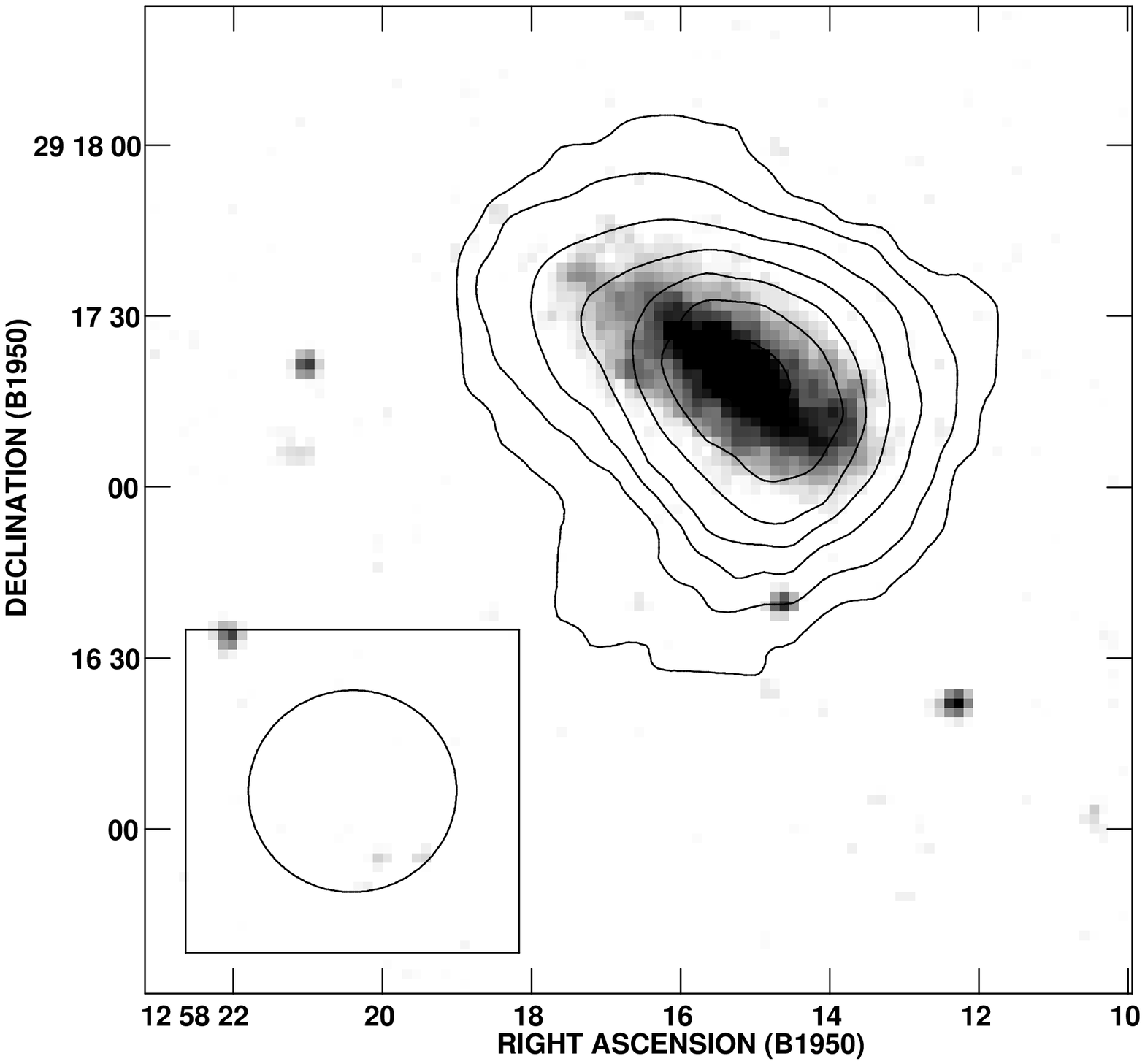}
      \caption{ H{\sc i} density distribution of
        IC~842 superposed on a DSS B-band gray scale image. The contours are
        0.3 (2.5 $\sigma$), 1.2, 2.5, 3.7, 4.9, 6.2, and $7.4 \times 10^{20}$
        cm$^{-2}$. The FWHM is indicated by the circle, $36.6^{\prime\prime}
        \times 35.4^{\prime\prime} $.
              }
    \vspace{3cm}
       \label{Fig}
   \end{figure}
%

   \begin{figure}
   \centering
   \includegraphics[width=7cm]{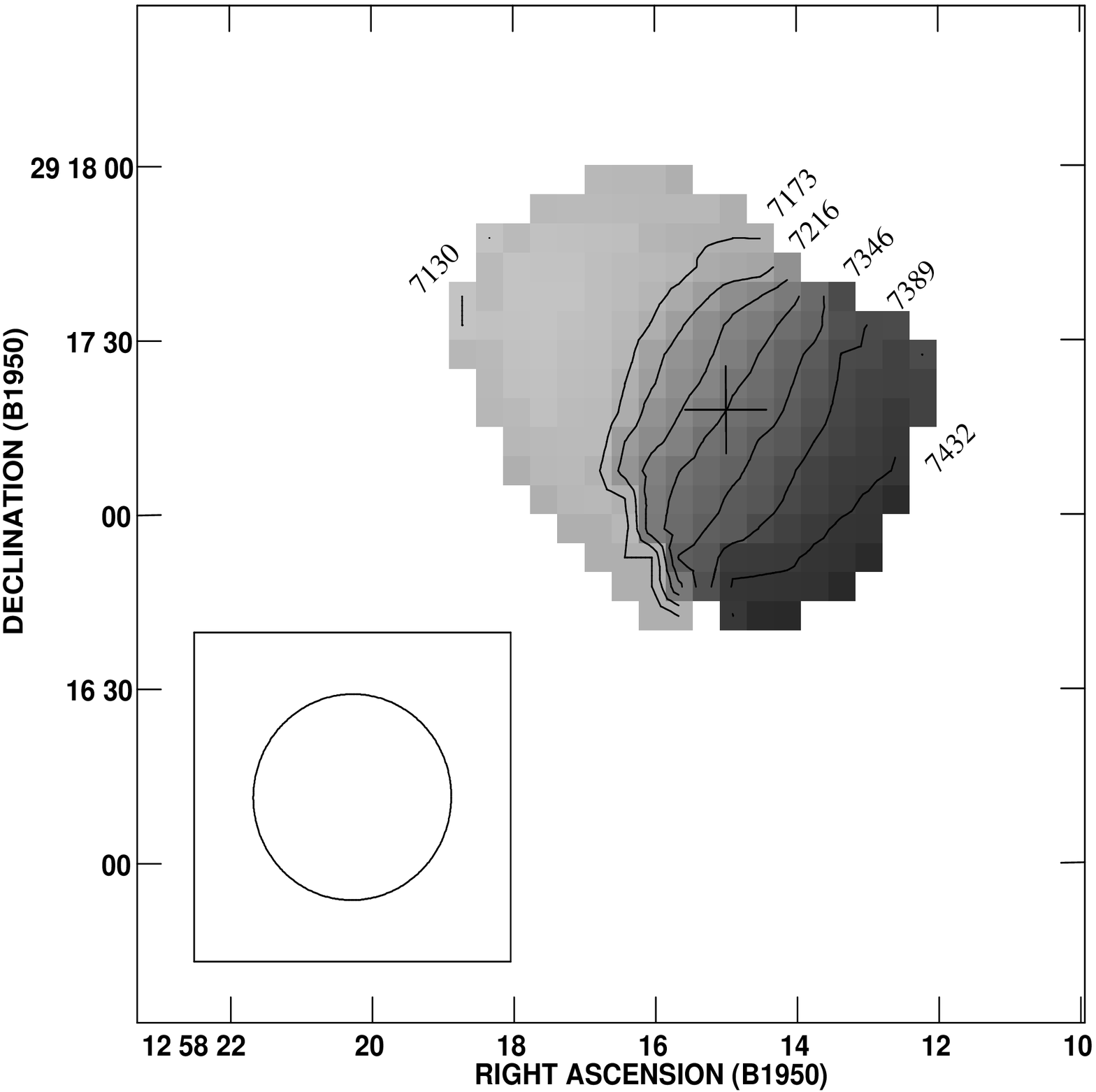}
      \caption{ Intensity weighted mean velocity
        field of IC~842.  The optical center of the galaxy is indicated with
        a cross.  The numbers indicate heliocentric velocity in \km.
        The FWHM is indicated by the circle, $36.6^ {\prime\prime}
        \times 35.4^{\prime\prime}$.  } \label{Fig} \end{figure}
%

\newpage

   \begin{figure}
   \centering
   \includegraphics[width=7cm]{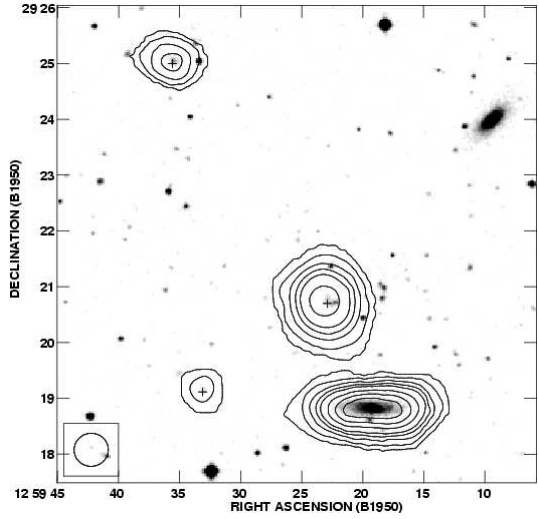}
      \caption{ H{\sc i} density distribution of
        IC~4088 and 3 neighbor dwarf systems, superposed on a DSS B-band gray
        scale image. The contours are 0.2 (2.5 $\sigma$), 1.0, 1.9, 2.9, 3.9,
        5.8, 7.8, and $9.7 \times 10^{20}$ cm$^{-2}$. The FWHM is indicated
        by the circle, $36.6^{\prime\prime} \times 35.4^{\prime\prime} $.
              }
    \vspace{3cm}
       \label{Fig}
   \end{figure}
%

   \begin{figure} \centering \includegraphics[width=7cm]{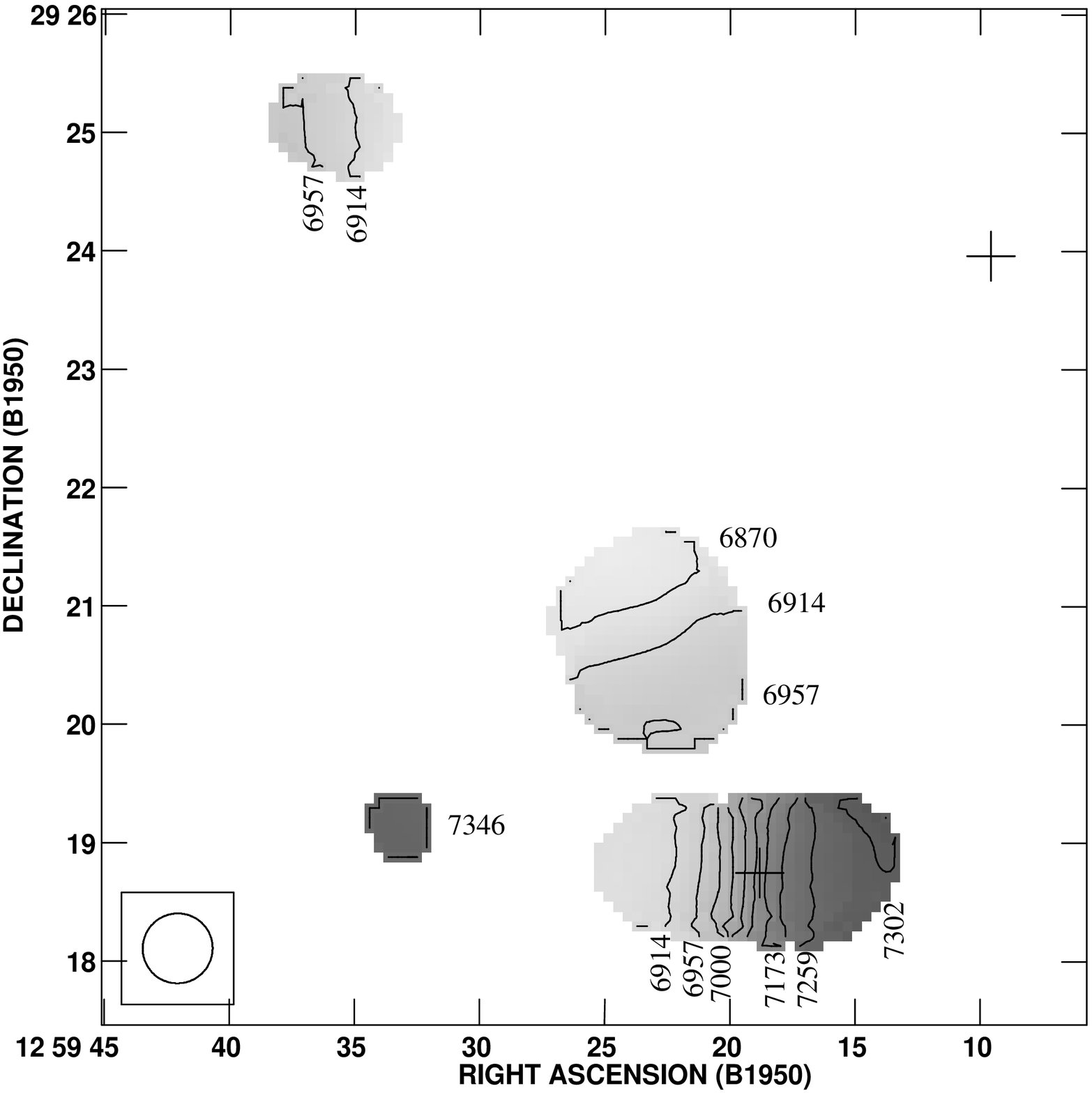}
   \caption{ Intensity weighted mean velocity field
   of IC~4088.  The optical center of the galaxies are indicated with
   crosses.  The numbers indicate heliocentric velocity in \km.  The FWHM is
   indicated by the circle, $36.6^ {\prime\prime} \times
   35.4^{\prime\prime}$.  } \label{Fig} \end{figure}
%

\newpage

   \begin{figure}
   \centering
   \includegraphics[width=7cm]{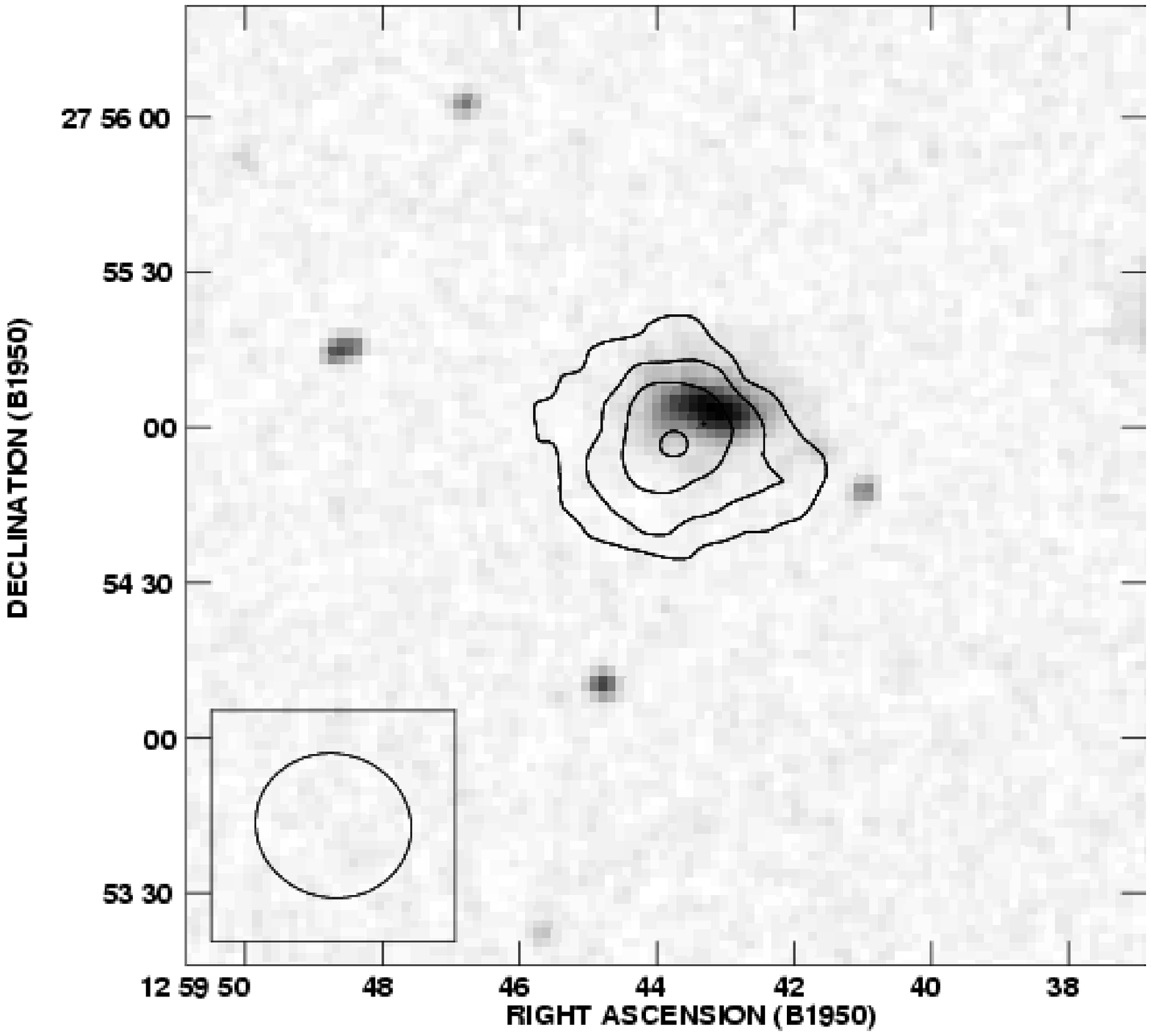}
      \caption{ H{\sc i} density distribution of
        NGC~4926--A, superposed on a DSS B-band gray scale image. The
        contours are 0.3 (2.5 $\sigma$), 1.4, and $2.1 \times 10^{20}$
        cm$^{-2}$. The FWHM is indicated by the circle, $35.2^{\prime\prime}
        \times 33.0 ^{\prime\prime} $.
              }
\vspace{15cm}
         \label{Fig}
   \end{figure}
%

\end{document}